\documentclass[letterpaper,preprint]{sig-alternate-10pt}
\usepackage{amsfonts}
\usepackage{amsfonts}
 \usepackage{amsfonts}
\usepackage{color}
\usepackage{pifont}
\usepackage{leading}
\usepackage{psfrag}
\usepackage{fancyhdr}
\newcommand{\bcap} {\hspace{2pt} \mathlarger{\cap}
\hspace{2pt}}
\newcommand{\h}{it holds that }
\newcommand{\la}{\label}
\newcommand{\f}{it follows that }
\newcommand{\n}{for all $n$ sufficiently large }

\newcommand{\bcup} {\hspace{2pt} \mathlarger{\cup}
\hspace{2pt}}
\newcommand{\qedb}{\hfill\ensuremath{\square}}
\usepackage[font=normalsize,labelfont=normalsize]{caption}
\usepackage{makecell}
\usepackage{booktabs}

\newfont{\secfntx}{ptmb8t at 11pt}
\newfont{\secfnty}{ptmb8t at 11.3pt}
\usepackage{url}
\usepackage{graphicx}
\usepackage{ntheorem}
 \usepackage{mathrsfs}
\usepackage{cite}
 \newfont{\secfnta}{ptmb8t at 12pt}

\usepackage[top=.75in, left=0.75in, right=0.75in, bottom=1in]{geometry}
\usepackage{relsize}

\def\centerhack#1{\hbox to 0pt{\hss\footnotesize #1\hss}}
\def\centerhackn#1{\hbox to 0pt{\hss #1\hss}}
\def\dchack#1{\vbox to 0pt{\vss{\hbox to 0pt{\hss#1\hss}}\vss}}

    \makeatletter
    \def\@listi{\leftmargin\leftmargini
        \parsep 1\p@ \@plus0\p@ \@minus\p@
        \topsep 2\p@   \@plus0\p@ \@minus\p@
        \itemsep1\p@ \@plus0\p@ \@minus\p@}
    \let\@listI\@listi\@listi
    \makeatother

\newcommand*\mcapinn[2]{\vcenter{\hbox{$\mathsurround=0pt
\ifx\displaystyle#1\textstyle\else#1\fi\bigcap$}}}

\newcommand*\mcupinn[2]{\vcenter{\hbox{$
\bigcup$}}}

\usepackage{cases}

\newtheorem{lem}{Lemma}
\newtheorem{thm}{Theorem}

\newtheorem{rem}{Remark}
\newtheorem{cor}{Corollary}

\newcommand*{\refname}{Bibliography}

\begin{document}

\pagestyle{fancy} 
\fancyhead[L]{Annual Allerton Conference on Communication, Control, and
  Computing (Allerton) 2014\\\small\url{http://dx.doi.org/10.1109/ALLERTON.2014.7028605}\\\small\url{http://ieeexplore.ieee.org/xpls/abs_all.jsp?arnumber=7028605}}
 
\fancyhead[R]{~}

%

\title{Connectivity in Secure Wireless Sensor Networks\\under
Transmission Constraints}



\numberofauthors{3} 
%
\author{
\alignauthor
Jun Zhao\\
       \affaddr{CyLab and Dept.
of ECE}\\
       \affaddr{Carnegie Mellon University}\\
       \affaddr{Pittsburgh, PA 15213}\\
       \email{junzhao@cmu.edu}
\alignauthor Osman Ya\u{g}an\\
       \affaddr{CyLab and Dept.
of ECE}\\
       \affaddr{Carnegie Mellon University}\\
       \affaddr{Moffett Field, CA 94035}\\
       \email{oyagan@ece.cmu.edu}
\alignauthor Virgil Gligor\\
       \affaddr{CyLab and Dept.
of ECE}\\
       \affaddr{Carnegie Mellon University}\\
       \affaddr{Pittsburgh, PA 15213}\\
       \email{gligor@cmu.edu}}


\markboth{a}{b}

\maketitle

%
%

\fontsize{10}{12}

\selectfont

\begin{abstract}

In wireless sensor networks (WSNs), the Eschenauer--Gligor (EG) key
pre-distribution scheme is a widely recognized way to secure
communications. Although the connectivity properties of secure WSNs
with the EG scheme have been extensively investigated, few results
address physical transmission constraints. These constraints reflect
real--world implementations of WSNs in which two sensors have to be
within a certain distance from each other to communicate. In this
paper, we present the first \emph{zero--one laws} for connectivity
in WSNs employing the EG scheme under transmission constraints.
These laws improve recent results \cite{ISIT_RKGRGG,Krzywdzi}
significantly, are sharp, and help specify the \emph{critical
transmission ranges} for connectivity. Our analytical findings,
which are also confirmed via numerical experiments, provide precise
guidelines for the design of secure WSNs in practice. The
application of our theoretical results to frequency hopping of
wireless networks is discussed in some detail.


\end{abstract}

 \vspace{-1pt} \terms{Theory} \vspace{-1pt}

\vspace{-1pt}


\section{INTRODUCTION} \label{introduction}

The Eschenauer--Gligor key pre-distribution scheme \cite{virgil} is
regarded as a typical approach to secure communications in wireless
sensor networks (WSNs). In this scheme (referred to as the EG scheme
hereafter), each sensor is independently assigned the same number of
distinct cryptographic keys selected uniformly at random from a key
pool before deployment. After deployment, any two sensors establish
a link between them, if they share at least one key.

Connectivity in secure WSNs employing the EG scheme has been
extensively studied in the literature
\cite{r1,zz,ryb3,yagan,ISIT_RKGRGG,Krzywdzi}. However, most existing
research \cite{r1,zz,ryb3,yagan} unrealistically assumes
unconstrained sensor-to-sensor communications; i.e., any two sensors
can communicate regardless of the distance between them.  Only two
recent results \cite{ISIT_RKGRGG,Krzywdzi} take transmission
constraints into consideration, but do not provide zero--one laws
for connectivity.\vspace{-.06pt}

In this paper, we establish the first and also sharp \emph{zero--one
laws} for connectivity in WSNs using the EG scheme under practical
transmission constraints. We present significantly improved
 conditions for asymptotic connectivity over those of Krishnan
\emph{et al.} \cite{ISIT_RKGRGG} and Krzywdzi\'{n}ski and Rybarczyk
\cite{Krzywdzi}, and also demonstrate that as the parameters move
further away from these conditions, the network rapidly becomes
asymptotically disconnected. Our results provide useful guidelines
for dimensioning the EG scheme and adjusting sensor transmission
power to ensure network connectivity. Moreover, our zero--one laws
enable us to determine the \emph{critical transmission ranges} for
connectivity. Intuitively, as the transmission range surpasses
(resp., falls below) the critical value and grows (resp., declines)
further, the network immediately enters an asymptotically connected
(resp., disconnected) state.


%

To model transmission \vspace{-.01pt}constraints, we use the popular
\emph{disk model}
\cite{citeulike:505396,Gupta98criticalpower,penrose,Li:2003:FTD:778415.778431,wanAsymptoticCritical},
in which \vspace{-.01pt}each sensor's transmission area is a disk
with a uniform distance\vspace{-.01pt} as its radius; i.e.,  two
sensors have to be within\vspace{-.01pt} the radius distance to
communicye directly. The network area\vspace{-.01pt} in our analysis
is either a torus or a square. The square\vspace{-.01pt} accounts
for the real--world \emph{boundary effect}\vspace{-.01pt} whereby
some transmission region of a sensor close\vspace{-.01pt} to the
network boundary may falls outside\vspace{-.01pt} the network field.
In contrast, the torus eliminates the\vspace{-.01pt} boundary
effect.

%

The rest of the paper is organized as follows.\vspace{-.01pt} In
Section \ref{sec:NetworkModel}, we describe the system
model.\vspace{-.01pt} Section \ref{sec:main:res}
presents\vspace{-.01pt} the main results, leading to the discussion
of critical transmission ranges in \vspace{-.01pt}Section
\ref{sec:ctr}. Afterwards, we explain the practicality of theorems'
conditions \vspace{-.01pt} in Section \ref{sec:pra:con}. We provide
numerical experiments\vspace{-.01pt} in Section \ref{sec:expe}. In
Section \ref{sec:disu:Mobility:fh}, we discuss\vspace{-.01pt} the
application of our results to frequency hopping in detail. Section
\ref{related}\vspace{-.01pt} reviews related work; and Section
\ref{sec:Conclusion} concludes the paper. The Appendix contains the
zero--law proofs and the sketch of the (simpler) one--law proofs.

\section{SYSTEM MODEL}
\label{sec:NetworkModel}

In a WSN with size $n$ and sensor set $\mathcal
{V}\hspace{-2pt}=\hspace{-2pt}\{v_1,\hspace{-1.27pt} v_2,
\hspace{-1.27pt}\ldots\hspace{-1pt}, \hspace{-1.27pt}v_n\}$, the EG
scheme independently assigns a set of $K_n$ distinct cryptographic
keys, which are selected {uniformly at random} from a pool of $P_n$
keys, to each sensor node. The set of keys of each sensor is called
the key ring and is denoted by $S_x$ for sensor $v_x$. The EG scheme
is modeled by a \emph{random key graph}
\cite{ryb3,yagan,ISIT_RKGRGG}, denoted by $G_{RKG}(n, K_n, P_n)$ in
which an edge exists between two nodes\footnote{The terms sensor and
node are interchangeable.} $v_x$ and $v_y$ if and only if they
possess at least one common key; i.e., the event $[S_x \bcap S_y
\neq \emptyset]$, denoted by $K_{xy}$, holds. As for the sensor
distribution, we consider that the $n$ nodes
 are independently and uniformly deployed in a network area $\mathcal {A}$.
The disk model induces a \emph{random geometric graph}
\cite{citeulike:505396,ISIT_RKGRGG,Krzywdzi,penrose,Gupta98criticalpower,wanAsymptoticCritical},
denoted by $G_{RGG}(n,r_n,\mathcal {A})$, in which an edge
  exists between two sensors if and only if their distance is no
   greater than $r_n$. In a secure WSN using the EG scheme under the disk
model, two sensors $v_x$ and $v_y$ establish a direct link between
them if and only if they share at least one key and are within
distance $r_n$. We denote the event establishing this direct link by
$E_{xy}$. If we let graph $G(n, \theta_n, \mathcal{A})$ model such a
WSN, it is straightforward to see $G(n, \theta_n, \mathcal{A})$ is
the intersection of random key graph $G_{RKG}(n, K_n, P_n)$ and
random geometric graph $G_{RGG}(n,r_n,\mathcal {A})$; namely,
\begin{align}
G(n, \theta_n, \mathcal{A}) &  =G_{RKG}(n, K_n, P_n) \bcap
G_{RGG}(n,r_n,\mathcal {A}),\nonumber
\end{align}
 where parameters $K_n, P_n$ and $r_n$ are together represented by
$\theta_n$. Also, if we let region $\mathcal {A}$ be either a torus
$\mathcal {T}$ or a square $\mathcal {S}$, each with a unit
area, we obtain the two graphs 
\begin{align}
G(n, \theta_n, \mathcal{T}) &  =G_{RKG}(n, K_n, P_n) \bcap
G_{RGG}(n,r_n,\mathcal {T}), \nonumber
\end{align}
and
\begin{align}
G(n, \theta_n, \mathcal{S}) &  =G_{RKG}(n, K_n, P_n) \bcap
G_{RGG}(n,r_n,\mathcal {S}). \nonumber
\end{align}

We let $p_s$ be the probability of key sharing between two sensors
and note that $p_s$ is also the edge probability in random key graph
$G_{RKG}(n, K_n, P_n)$. It holds that $p_{s} = \mathbb{P} [{K}_{xy}
] = \mathbb{P}[S_x \cap S_y \neq \emptyset]$. Clearly, if $P_n < 2
K_n$, then $p_{s} = 1$. If $P_n \geq 2 K_n$,
 as shown in previous work \cite{r1,ryb3,yagan}, we have $p_{s} = 1- {\binom{P_n- K_n}{K_n}
 }\big/
{\binom{P_n}{K_n}}$. \hspace{-1pt}If $P_n \geq 2K_n $, by
{\cite[Lemma 6]{bloznelis2013}}, it further holds that
\begin{align}
p_{s}   & \leq {{K_n}^2}/{P_n} \label{psleqKnPn}.
\end{align}
By {\cite[Lemma 8]{ZhaoYaganGligor}}, (\ref{psleqKnPn}) implies that
 if ${{K_n}^2}/{P_n}  =  o(1)$,
then
\begin{align}
p_{s} & = {{K_n}^2}/{P_n} \cdot
\big[1-O\big({{K_n}^2}/{P_n}\big)\big] \sim {{K_n}^2}/{P_n}.
\label{pssim}
\end{align}
We will frequently use (\ref{psleqKnPn}) and (\ref{pssim})
throughout the paper.\footnote{We use the standard Landau asymptotic
notation $o(\cdot), O(\cdot), \omega(\cdot),
\Omega(\cdot),\Theta(\cdot)$ and $ \sim$; in particular, for two
positive functions $f_1(n)$ and $f_2(n)$, the relation $f_1(n) \sim
f_2(n)$ means $\lim_{n \to
  \infty} {f_1(n)}/{f_2(n)}=1$.}

Let $p_{e}$ be the probability that a link exists between two
sensors in the WSN modeled by graph $G(n, \theta_n, \mathcal{A})$;
i.e., $p_{e}$ is the edge probability in $G(n, \theta_n,
\mathcal{A})$. It holds that $p_{e} = \mathbb{P}[E_{xy}]$. When
$\mathcal{A}$ is the torus $\mathcal{T}$, clearly $p_{e}$ equals
$\pi {r_n}^2 \cdot p_{s} $; and if ${{K_n}^2}/{P_n} = o(1)$, then
$p_{e} \sim \pi {r_n}^2 \cdot {{K_n}^2}/{P_n}$ by (\ref{pssim}).
When $\mathcal{A}$ is the square $\mathcal{S}$, it is a simple
matter to show $p_{e} \leq \pi {r_n}^2 \cdot p_{s} $ and $p_{e} \geq
(1 - 2 {r_n})^2 \cdot \pi {r_n}^2 \cdot p_{s} $, yielding $p_{e}
\sim \pi {r_n}^2 \cdot p_{s} $ if $r_n = o(1)$. Therefore, on
$\mathcal{S}$, if $r_n = o(1)$ and ${{K_n}^2}/{P_n} = o(1)$, we
further obtain $p_{e} \sim \pi {r_n}^2 \cdot {{K_n}^2}/{P_n}$ in
view of (\ref{pssim}).


In addition to random key graphs and random geometric graphs, the
Erd\H{o}s-R\'enyi graph \cite{citeulike:4012374} has also been
extensively studied. An Erd\H{o}s--R\'{e}nyi graph $G_{ER}(n, p_n)$
is defined on a set of $n$ nodes such that any two nodes establish
an edge in between independently with probability $p_n$. As already
shown in the literature \cite{r1,ryb3,zz,yagan}, random key graph
$G_{RKG}(n, K_n, P_n)$ and Erd\H{o}s-R\'enyi graph $G_{ER}(n,p_n)$
have similar connectivity properties when they are {\em matched}
through edge probabilities; i.e. when
$p_s\hspace{-3pt}=\hspace{-3pt}p_n$. Hence, it would be tempting to
exploit this analogy
and conclude that connectivity in $G(n, \theta_n, \mathcal{A})$
(i.e., $G_{RKG}(n, K_n, P_n) \bcap G_{RGG}(n,r_n,\mathcal {A})$) is
similar to that of $G_{ER}(n,p_n)\bcap G_{RGG}(n,r_n,\mathcal {A})$,
which was recently established~\cite{Penrose2013}. However tempting,
such heuristic approaches do not work as graphs $G_{ER}(n, p_s)$ and
$G_{RKG}(n,K_n,P_n)$ (and their respective intersections) are quite
different.
 For instance, in $G_{ER}(n, p_s)$, for any three nodes $v_x, v_y$ and
 $v_z$, the event that $v_x$ has edges with both $v_y$ and $v_z$, is
 independent of the event that $v_y$ and $v_z$ has an edge
 between them. However, in $G_{RKG}(n,K_n,P_n)$, these two events are not
 independent from each other, since the event that $v_x$ has edges
 with both $v_y$ and $v_z$ means that the key rings $S_y$ and $S_z$
 of $v_y$ and $v_z$ respectively both have intersections with the key
 ring $S_x$ of $v_x$. This has an impact on whether $S_y$ and $S_z$
 intersects.
In fact, it has been formally proven \cite{yagan,bloznelis2013} that
graphs $G_{ER}(n, p_s)$ and $G_{RKG}(n,K_n,P_n)$ exhibit different
characteristics in terms of properties including {\em clustering
coefficient, number of triangles}, etc.

\section{THE MAIN RESULTS} \label{sec:main:res}

We detail the main results below. The notation ``$\ln$'' stands for
the natural logarithm function.

%

\subsection{Connectivity in a Secure WSN on a Torus}

Theorem \ref{thm:t} presents a zero--one law for connectivity in
 $G(n, \theta_n, \mathcal{T})$, which models a secure WSN
working under the EG scheme and the disk model on a unit torus.

\begin{thm} \label{thm:t}
Let graph $G(n, \theta_n, \mathcal{T})$ be the intersection of
random key graph $G_{RKG}(n, K_n, P_n)$ and random geometric graph
$G_{RGG}(n, r_n, \mathcal{T})$ on a unit torus $\mathcal{T}$, where
there exist some $\mu_n = \omega(1)$ and constant $c_1$ such that
\begin{align}
\max\bigg\{ \frac{\ln n}{\ln \ln n}, \hspace{1.5pt} \mu_n \cdot
\sqrt{\frac{P_n \ln n}{n}}
 \bigg\} &  \leq   K_n  \leq   c_1 \sqrt{\frac{P_n}{\ln n}}
\label{thm:t_Kn}
\end{align}
for all $n$ sufficiently large. For all $n$, let the sequence
$\alpha_n$  be defined through
\begin{align}
 \pi {r_n}^2 \cdot \frac{{K_n}^2}{P_n} & = \frac{\ln n +
 \alpha_{n}}{n}. \label{thm:t:rnKnPn}
\end{align}
Then
\begin{align}
  \lim\limits_{n \to \infty}\mathbb{P}\left[ \begin{array}{l} \textrm{$G(n, \theta_n, \mathcal{T})$}
 \\  \textrm{is connected.}
\end{array} \right] & =
\begin{cases} 0, ~~\textrm{if $\lim\limits_{n \to \infty}{\alpha_n}
=-\infty$}, \\  1, ~~\textrm{if $\lim\limits_{n \to
\infty}{\alpha_n} =\infty$}.\end{cases} \nonumber
\end{align}
\end{thm}
\begin{rem} \label{rem:t:edge}
Under (\ref{thm:t_Kn}), we obtain $\frac{{K_n}^2}{P_n} \leq
\frac{{c_1}^2}{\ln n} = o(1)$.
Then as noted in Section \ref{sec:NetworkModel}, %
 $\pi {r_n}^2 \cdot
\frac{{K_n}^2}{P_n}$ in (\ref{thm:t:rnKnPn}) asymptotically equals
the edge probability in graph $G(n, \theta_n, \mathcal{T})$.
\end{rem}

\subsection{Connectivity in a Secure WSN on a Square}

Theorem \ref{thm:s} gives a zero--one law for connectivity in $G(n,
\theta_n, \mathcal{S})$, which models a secure WSN working under the
EG scheme and the disk model on a unit
 square.

\begin{thm} \label{thm:s}
Let graph $G(n, \theta_n, \mathcal{S})$ be the intersection of
random key graph $G_{RKG}(n, K_n, P_n)$ and random geometric graph
$G_{RGG}(n, r_n, \mathcal{S})$ on a unit square $\mathcal{S}$, where
there exist some constants $c_2
>0,$ $0<c_3<1$, $c_4>0$ and $\nu_n = o(1)$
 such that
\begin{align}
 c_2 \sqrt{\frac{P_n \ln n}{n^{c_3}}} & \leq K_n \leq
\min\Bigg\{\nu_n \cdot  \sqrt{\frac{P_n}{\ln n}},
\hspace{1.5pt}{\frac{ c_4 P_n}{n\ln n}}\hspace{2pt}\Bigg\}
\label{thm:s_Kn}
\end{align}
for all $n$ sufficiently large. Assume that $\frac{{K_n}^2}{P_n}
\cdot n^{1/3}\ln n$ \emph{either} is bounded for all $n$ \emph{or}
converges to $\infty$ as $n \to \infty$, and for all $n$ let the
sequence $\alpha_n$ be defined through
\begin{align}
\hspace{-117pt} \pi {r_n}^2 \cdot  \textrm{\fontsize{9.5}{11}
\selectfont $\frac{{K_n}^2}{P_n}$} & \nonumber
\end{align}\vspace{-10pt}
\begin{align}
&\hspace{-4pt}=\begin{cases} \hspace{-4pt}\textrm{\fontsize{13}{15}
\selectfont $\frac{\ln \frac{n P_{n}}{{K_{n}}^2}
\hspace{2pt}-\hspace{2pt} \ln \ln \frac{n P_{n}}{{K_{n}}^2}
\hspace{2pt}+\hspace{2pt} \alpha_{n} }{n}$} ,
&\hspace{-6pt}\textrm{for~} \textrm{\large \selectfont
$\frac{{K_n}^2}{P_n}$} = \omega
\textrm{\large \selectfont $\left( \frac{1}{n^{1/3}\ln n} \right)$}, \\
\vspace{-7pt} \\
 \hspace{-4pt}\textrm{\fontsize{13}{15} \selectfont $\frac{ 4\ln \frac{P_n}{{K_n}^2} \hspace{2pt}
 -\hspace{2pt} 4\ln \ln \frac{P_n}{{K_n}^2} \hspace{2pt}+\hspace{2pt}
\alpha_{n}  }{n}$} , &\hspace{-6pt}\textrm{for~} \textrm{\large
\selectfont $\frac{{K_{n}}^2}{P_{n}}$}  = O \textrm{\large
\selectfont $\left( \frac{1}{n^{1/3}\ln n} \right)$}.\end{cases}
\label{thm:s:rnKnPn}
\end{align}
\normalsize \selectfont Then
\begin{align}
  \lim\limits_{n \to \infty}\mathbb{P}\left[ \begin{array}{l} \textrm{$G(n, \theta_n, \mathcal{S})$}
 \\  \textrm{is connected.}
\end{array} \right] & =
\begin{cases} 0, ~~\textrm{if $\lim\limits_{n \to \infty}{\alpha_n}
=-\infty$}, \\  1, ~~\textrm{if $\lim\limits_{n \to
\infty}{\alpha_n} =\infty$.}\end{cases} \nonumber
\end{align}
\end{thm}

\begin{rem}  \label{rem:s:edge}
Under (\ref{thm:s_Kn}), we have $\frac{{K_n}^2}{P_n} \leq
\frac{{\nu_n}^2}{\ln n} = o(1)$. If it further holds $r_n = o(1)$
(this is true with confined $\alpha_n$), then as established in
Section
\ref{sec:NetworkModel}, 
 $\pi {r_n}^2
\cdot \frac{{K_n}^2}{P_n}$ in (\ref{thm:s:rnKnPn}) asymptotically
equals the edge probability in graph $G(n, \theta_n, \mathcal{S})$.
\end{rem}

\newpage

\section{CRITICAL TRANSMISSION RANGES}\label{sec:ctr}

\subsection{The Critical Transmission Range
 for Connectivity in a Secure WSN on a Unit Torus}

By Theorem \ref{thm:t}, under condition (\ref{thm:t_Kn}), we can
determine the critical transmission range $r_n^{*}(\mathcal{T})$ for
connectivity in a secure WSN on a unit torus modeled by graph $G(n,
\theta_n, \mathcal{T})$ through
\begin{align}
\pi \big[{r_n^{*}(\mathcal{T})}\big]^2 \cdot \frac{{K_n}^2}{P_n} & =
\frac{\ln n }{n}, \nonumber
\end{align}
inducing the following expression of $r_n^{*}(\mathcal{T})$:
\begin{align}
r_n^{*}(\mathcal{T}) = \sqrt{\frac{\ln n}{\pi n} \cdot
\frac{P_n}{{K_n}^2}}. \label{ctr:t}
\end{align}
By (\ref{ctr:t}), it is clear that with $n$ fixed,
$r_n^{*}(\mathcal{T})$ decreases as $\frac{{K_n}^2}{P_n}$ increases.
This is expected since as mentioned in Remark \ref{rem:t:edge} after
Theorem \ref{thm:t},
$\frac{{K_n}^2}{P_n}$
asymptotically equals the probability that two sensors share at
least one key; and $\pi {r_n}^2 \cdot \frac{{K_n}^2}{P_n}$
asymptotically equals the edge probability in $G(n, \theta_n,
\mathcal{T})$. As the probability of key sharing increases, sensors
can reduce their transmission ranges to maintain network
connectivity.

We explain $r_n^{*}(\mathcal{T}) = o(1)$, which is anticipated as
the node density $n$ grows to $\infty$. From $K_n \geq
\sqrt{\frac{P_n \ln n}{n}} \cdot \mu_n$ in (\ref{thm:t_Kn}), where
$\mu_n = \omega(1)$, \f $\frac{{K_n}^2}{P_n} = \omega\big(\frac{\ln
n }{n}\big)$, which along with (\ref{ctr:t}) leads to
$r_n^{*}(\mathcal{T}) = o(1)$.

 \subsection{The Critical Transmission Range
 for Connectivity in a Secure WSN on a Unit Square} \label{s:ctr}

By Theorem \ref{thm:s}, under condition (\ref{thm:s_Kn}), we can
determine the critical transmission range $r_n^{*}(\mathcal{S})$ for
connectivity in a secure WSN on a unit square modeled by graph $G(n,
\theta_n, \mathcal{S})$ through
\begin{align}\nonumber
\hspace{-117pt} \pi \big[{r_n^{*}(\mathcal{S})}\big]^2  \cdot
\frac{{K_n}^2}{P_n}  & \nonumber
\end{align}\vspace{-7pt}
\begin{align}
&\hspace{-4pt}=\begin{cases} \hspace{-4pt}\textrm{\fontsize{13}{15}
\selectfont $\frac{\ln \frac{n P_{n}}{{K_{n}}^2}
\hspace{2pt}-\hspace{2pt} \ln \ln \frac{n P_{n}}{{K_{n}}^2}
\hspace{2pt} }{n}$} , &\hspace{-4pt}\textrm{for~} \textrm{\large
\selectfont $\frac{{K_n}^2}{P_n}$} = \omega
\textrm{\large \selectfont $\left( \frac{1}{n^{1/3}\ln n} \right)$}, \\
\vspace{-7pt} \\
 \hspace{-4pt}\textrm{\fontsize{13}{15}
 \selectfont $\frac{ 4\ln \frac{P_n}{{K_n}^2} \hspace{2pt}
 -\hspace{2pt} 4\ln \ln \frac{P_n}{{K_n}^2}
  \hspace{2pt} }{n}$} , &\hspace{-4pt}\textrm{for~} \textrm{\large \selectfont
$\frac{{K_{n}}^2}{P_{n}}$}  = O \textrm{\large \selectfont $\left(
\frac{1}{n^{1/3}\ln n} \right)$},\end{cases}\label{eq_pirn}
\end{align}
so $r_n^{*}(\mathcal{S})$ is specified by
\begin{align}
&\hspace{-1pt}r_n^{*}(\mathcal{S})\hspace{-1pt}=\hspace{-1pt}\begin{cases}\hspace{-4pt}
\textrm{\fontsize{13}{15} \selectfont$\sqrt{\frac{\ln \frac{n
P_n}{{K_n}^2} \hspace{1.5pt}-\hspace{1.5pt} \ln \ln \frac{n
P_n}{{K_n}^2} }{\frac{\pi n
{K_n}^2}{P_n}}}$},&\hspace{-7.5pt}\textrm{for~} \textrm{\large
\selectfont $\frac{{K_n}^2}{P_n}$} \hspace{-1pt}=\hspace{-1pt}
\omega \textrm{\large \selectfont $\left(
\frac{1}{n^{1/3}\ln n} \right)$},  \\ \vspace{-7pt} \\
\hspace{-1pt} 2\hspace{-2pt}\textrm{\fontsize{13}{15}
\selectfont$\sqrt{\frac{ \ln \frac{P_n}{{K_n}^2}
\hspace{1.5pt}-\hspace{1.5pt} \ln \ln \frac{P_n}{{K_n}^2}
}{\frac{\pi n {K_n}^2}{P_n}}}$},&\hspace{-7.5pt}\textrm{for~}
\textrm{\large \selectfont $\frac{{K_n}^2}{P_n}$}
\hspace{-1pt}=\hspace{-1pt} O \textrm{\large \selectfont $\left(
\frac{1}{n^{1/3}\ln n} \right)$}.
\end{cases}  \label{ctr:s}
\end{align}
%

First, we show that for fixed and sufficiently large $n$, the
critical transmission range $r_n^{*}(\mathcal{S})$ decreases as
$\frac{{K_n}^2}{P_n}$ increases. Similar to the discussion on
$r_n^{*}(\mathcal{T})$, this is also expected in that as the
probability $\frac{{K_n}^2}{P_n}$ of key sharing increases, sensors
can reduce their transmission ranges to maintain network
connectivity. For the formal argument we note that $f(x) = \ln x -
\ln \ln x$ is an increasing function of $x$ for $x>e$ since its
derivative $f'(x) =\frac{1}{x}(1-\frac{1}{\ln x})$ is positive,
where $e$ is the base of $\ln$. From condition (\ref{thm:s_Kn}), we
have $\frac{ P_n}{{K_n}^2} \geq
 \frac{\ln n}{{\nu_n}^2}$, which implies that
$\frac{P_n}{{K_n}^2} >e$ for all $n$ sufficiently large. Hence, for
fixed and sufficiently large $n$,
 $\big(\ln \frac{P_n}{{K_n}^2}
 - \ln \ln \frac{P_n}{{K_n}^2}\big)$ and $\big(\ln \frac{n
P_n}{{K_n}^2} - \ln \ln \frac{n P_n}{{K_n}^2}\big)$ are both
increasing as $\frac{P_n}{{K_n}^2}$ increases, and hence decreasing
as $\frac{{K_n}^2}{P_n}$ increases. Hence, $r_n^{*}(\mathcal{S})$
also decreases as $\frac{{K_n}^2}{P_n}$ increases by (\ref{ctr:s}).

Second, we show that the critical transmission range
$r_n^{*}(\mathcal{S}) = o(1)$, which is anticipated as the node
density $n$ grows to $\infty$. From condition (\ref{thm:s_Kn}), for
all $n$ sufficiently large, we have $\frac{{c_2}^2\ln n}{n^{c_3}}
\leq \frac{{K_n}^2}{P_n} < \frac{{c_4}^2}{\ln n}$, or  $\frac{n\ln
n}{{c_4}^2} < \frac{nP_n}{{K_n}^2}
 \leq \frac{n^{c_3+1}}{{c_2}^2\ln n}$.
This implies that in the case $\frac{{K_n}^2}{P_n} = \omega\left(
\frac{1}{n^{1/3}\ln n} \right)$ of (\ref{eq_pirn}), the term $\pi
[{r_n^{*}(\mathcal{S})}]^2 \cdot \frac{{K_n}^2}{P_n}$ is
$\Theta\big(\frac{\ln n}{n}\big)$. The second case of
(\ref{eq_pirn}), namely $\frac{{K_n}^2}{P_n} = O \left(
\frac{1}{n^{1/3}\ln n} \right)$ or $ \frac{P_n}{{K_n}^2} = \Omega
\left( n^{1/3}\ln n\right)$, and condition
 $\frac{P_n}{{K_n}^2} \leq \frac{n^{c_3}}{{c_2}^2\ln n}$
derived from (\ref{thm:s_Kn}) imply that term $\pi
[{r_n^{*}(\mathcal{S})}]^2 \cdot \frac{{K_n}^2}{P_n}$ of
(\ref{eq_pirn}) is also $\Theta\big(\frac{\ln n}{n}\big)$. Hence,
the term $\pi [{r_n^{*}(\mathcal{S})}]^2 \cdot \frac{{K_n}^2}{P_n}$
is $\Theta\big(\frac{\ln n}{n}\big)$ in  (\ref{eq_pirn}). This fact
and condition $\frac{{K_n}^2}{P_n} \geq \frac{{c_2}^2\ln n}{n^{c_3}}
= \omega\big(\frac{\ln n}{n}\big)$ imply that $r_n^{*}(\mathcal{S})
= o(1)$.


Third, we relate the critical transmission ranges of the unit square
$\mathcal{S}$ and torus $\mathcal{T}$, namely $r_n^{*}(\mathcal{S})
\geq r_n^{*}(\mathcal{T})$ for all $n$ sufficiently large.
Intuitively, this relationships is caused by the boundary effects of
$\mathcal{S}$. Specifically, two sensors close to opposite edges of
the square may be unable to establish a link on the square
$\mathcal{S}$ but may have a link in between on the torus
$\mathcal{T}$ because of possible wrap-around connections on the
torus. In view of (\ref{ctr:t}) and (\ref{ctr:s}), to prove
$r_n^{*}(\mathcal{S}) \geq r_n^{*}(\mathcal{T})$, we only need to
show for all $n$ sufficiently large that \\
(i) $\big(\ln \frac{nP_n}{{K_n}^2} - \ln \ln \frac{n P_n}{{K_n}^2}
\geq \ln n\big)$ for $\frac{{K_n}^2}{P_n} = \omega \left(
\frac{1}{n^{1/3}\ln n} \right)$
and  \\
(ii) \hspace{-.5pt}$\big(4\ln \frac{P_n}{{K_n}^2} - 4\ln \ln \frac{
P_n}{{K_n}^2} \geq \ln n\big)$ \hspace{-.5pt}for
$\frac{{K_n}^2}{P_n} \hspace{-1pt}= \hspace{-1pt}O\left(
\frac{1}{n^{1/3}\ln n} \right)$. 

To prove (i), we recall that condition (\ref{thm:s_Kn}) implies
$\frac{ P_n}{{K_n}^2} \geq
 \frac{\ln n}{{\nu_n}^2}$ and $\frac{nP_n}{{K_n}^2}
 \leq  \frac{n^{c_3+1}}{{c_2}^2\ln
n}$, where $\nu_n = o(1)$. It follows that, for all $n$ sufficiently
large,
 $\frac{ P_n}{{K_n}^2} \geq
 \frac{\ln n}{{\nu_n}^2} \geq \ln \frac{n^{c_3+1}}{{c_2}^2\ln n} \geq \ln
\frac{nP_n}{{K_n}^2}$, implies $ \ln \frac{n P_n}{{K_n}^2} - \ln \ln
\frac{n P_n}{{K_n}^2} - \ln n = \ln \frac{ P_n}{{K_n}^2} - \ln \ln
\frac{nP_n}{{K_n}^2} \geq 0$, which proves (i).

To prove (ii), recall that $\frac{{K_n}^2}{P_n}  =
O\left(\frac{1}{n^{1/3}\ln n} \right)$ implies $\frac{P_n}{{K_n}^2}
= \Omega\left(n^{1/3}\ln n\right)$,
 and hence
$\frac{P_n}{{K_n}^2} \geq c_5 n^{1/3}\ln n $ for some constant $c_5
>0$. Together with $\frac{P_n}{{K_n}^2} \leq
\frac{n^{c_3}}{{c_2}^2\ln n}$, this implies that, for all $n$
sufficiently large, $4\ln \frac{P_n}{{K_n}^2} - 4\ln \ln \frac{
P_n}{{K_n}^2} \geq \ln n$, which proves (ii).

\subsection{Phase Transition in the Critical Range $r_n^{*}(\mathcal{S})$}

\begin{cor} \label{coro}
Under the conditions of {\bf Theorem 2}, a phase transition occurs
for $r_n^{*}(\mathcal{S})$ when $\frac{{K_n}^2}{P_n}$ is of the
order of $\frac{1}{n^{1/3}\ln n}$, namely
\begin{align}\nonumber
\hspace{-77pt} \lim\limits_{n \to \infty} \bigg\{\bigg[ \pi
\big[{r_n^{*}(\mathcal{S})}\big]^2 \cdot \frac{{K_n}^2}{P_n}\bigg]
\bigg/\bigg(\frac{\ln n}{n}\bigg)\bigg\}  & \nonumber
\end{align}\vspace{-10pt}
\begin{align}
&\hspace{-7pt}=\hspace{-2pt}\begin{cases} \hspace{-2pt} \textrm{$1
\hspace{-3pt}+\hspace{-3pt} \lim\limits_{n \to \infty}\hspace{-2pt}
\Big({\ln \frac{P_n}{{K_n}^2}}\Big/{\ln
n}\Big)$},&\hspace{-8pt}\textrm{for~} \textrm{\large \selectfont
$\frac{{K_n}^2}{P_n}$} \hspace{-1.7pt}=\hspace{-1.7pt} \omega
\textrm{\large \selectfont $\left( \frac{1}{n^{1/3}\ln n} \right)$},
\\ \vspace{-10pt} \\ \hspace{-2pt}
 \textrm{$4\lim\limits_{n \to \infty}\hspace{-2pt} \Big({\ln \frac{P_n}{{K_n}^2}}\Big/{\ln
n}\Big)$},&\hspace{-8pt}\textrm{for~} \textrm{\large \selectfont
$\frac{{K_n}^2}{P_n}$} \hspace{-1.7pt}=\hspace{-1.7pt} O
\textrm{\large \selectfont $\left( \frac{1}{n^{1/3}\ln n} \right)$}.
\end{cases} \label{d}
\end{align}
\end{cor}

To prove this corollary, we recall that $\frac{{K_n}^2}{P_n}  \geq
\frac{{c_1}^2\ln n}{n^{c_2}}$ in (\ref{thm:s_Kn}), which implies
that
\begin{align}
\ln \ln \frac{n P_{n}}{{K_{n}}^2}  &  \leq \ln \ln
\frac{n^{c_2+1}}{{c_1}^2\ln n} = o(\ln n) . \label{olnn}
\end{align}

For the case $\frac{{K_n}^2}{P_n} = \omega \left(
\frac{1}{n^{1/3}\ln n} \right)$ of (\ref{eq_pirn}) and given
(\ref{olnn}), if $\lim\limits_{n \to \infty} \Big({\ln
\frac{P_n}{{K_n}^2}}\Big/{\ln n}\Big) = a$, we obtain
\begin{align}
& \bigg\{ \pi \big[{r_n^{*}(\mathcal{S})}\big]^2 \cdot
\frac{{K_n}^2}{P_n}\bigg\} \bigg/\bigg(\frac{\ln n}{n}\bigg)
 \nonumber  \\ & \quad = \bigg({\ln \frac{n P_{n}}{{K_{n}}^2} - \ln \ln \frac{n
P_{n}}{{K_{n}}^2} }\bigg)\bigg/{\ln n} \nonumber  \\ & \quad \to a +
1,\textrm{ as }n \to \infty. \label{d1}
\end{align}
Condition (\ref{thm:s_Kn}) implies $\frac{P_n}{{K_n}^2}
 \geq \frac{\ln n}{{c_4}^2} $, which along with
$\frac{P_n}{{K_n}^2} = o \left( n^{1/3}\ln n \right)$ leads to $a
\in [0,\frac{1}{3}]$.
%

For the case $\frac{{K_n}^2}{P_n} = O \left( \frac{1}{n^{1/3}\ln n}
\right)$ of (\ref{eq_pirn}) and given (\ref{olnn}), if
$\lim\limits_{n \to \infty} \Big({\ln \frac{P_n}{{K_n}^2}}\Big/{\ln
n}\Big) = a$, we have
\begin{align}
& \bigg\{ \pi \big[{r_n^{*}(\mathcal{S})}\big]^2 \cdot
\frac{{K_n}^2}{P_n}\bigg\} \bigg/\bigg(\frac{\ln n}{n}\bigg)
 \nonumber  \\ & \quad = \bigg({4\ln \frac{P_{n}}{{K_{n}}^2}
  - 4\ln \ln \frac{
P_{n}}{{K_{n}}^2} }\bigg)\bigg/{\ln n} \nonumber  \\ & \quad \to 4
a,\textrm{ as }n \to \infty. \label{d2}
\end{align}
Condition (\ref{thm:s_Kn}) implies $\frac{P_n}{{K_n}^2}
 \leq \frac{n^{c_3}} {{c_2}^2\ln n}$ where $0<c_3<1$. This and
$\frac{P_n}{{K_n}^2} = \Omega \left( n^{1/3}\ln n\right)$ show
 $a \in [\frac{1}{3},c_3]$.%
%


Finally, (\ref{d1}) and (\ref{d2}) together yield (\ref{d}). To
understand $\lim\limits_{n \to \infty} \Big({\ln
\frac{P_n}{{K_n}^2}}\Big/{\ln n}\Big) = a$, it is a simple matter to
check that such condition means for arbitrary $\epsilon>0$,
\begin{align}
n^{-a-\epsilon} \leq \frac{{K_n}^2}{P_n} & \leq
n^{-a+\epsilon},\textrm{ for }n\textrm{ sufficiently large}.
\label{KnPnepsilon}
\end{align}
Example values of $\frac{{K_n}^2}{P_n}$ satisfying
(\ref{KnPnepsilon}) are $c_6n^{-a}(\ln n)^{c_7}$, $c_8 n^{-a}(\ln
n)^{c_9}(\ln n \ln n)^{c_{10}}$, where $c_6, c_7, c_8, c_9$ and
$c_{10}$ are all arbitrary constants (of course, the coefficients
$c_6$ and $c_8$ should be positive).

Corollary \ref{coro} enables us to compare our results with the best
known to date (viz., Section \ref{related}), where the upper bounds
on ${{\lim\limits_{n \to \infty} \big\{ \pi
\big[{r_n^{*}(\mathcal{S})}\big]^2 \cdot \frac{{K_n}^2}{P_n}
\big/\big(\frac{\ln n}{n}\big)\big\}}}$ are $8$ and $2\pi$,
respectively~\cite{Krzywdzi, ISIT_RKGRGG}. Note that phase
transitions are not observed for the critical range
$r_n^{*}(\mathcal{T})$ of a torus $\mathcal{T}$.

%
%

\section{PRACTICALITY}\label{sec:pra:con}

In practical implementations of WSNs, $K_n$ controls the number of
keys in each sensor's memory, and should be small \cite{virgil}
compared to both $n$ and $P_n$ due to limited memory and
computational
capability of sensors. 

\subsection{Practicality of the Theorem \ref{thm:t} Conditions}

By (\ref{thm:t_Kn}), we obtain constraints (i) $\frac{{K_n}^2}{P_n}
= \omega\big( \frac{\ln n}{n}\big)$ and (ii) $\frac{{K_n}^2}{P_n}  =
O\big(\frac{1}{\ln n}\big)$, in view of $\frac{{K_n}^2}{P_n} \geq
{\mu_n}^2 \cdot \frac{\ln n}{n}$ with $\mu_n = \omega(1)$, and
$\frac{{K_n}^2}{P_n}
 \leq \frac{{c_1}^2}{\ln n}$. In addition, (\ref{thm:t_Kn}) enforces
(iii) $K_n \geq \frac{\ln n}{\ln \ln n}$. All these contraints (i),
(ii) and (iii) hold in real--world WSN applications.
%
%
%

\subsection{Practicality of the Theorem \ref{thm:s} Conditions}



We first discuss the relationship enforced between $P_n$ and $K_n$
by (\ref{thm:s_Kn}). It is a simple matter to see (\ref{thm:s_Kn})
is equivalent
  to the combination of (iv) $\frac{K_n}{P_n} \leq \frac{c_4}{n \ln n}$\
   and (v) $\frac{{c_2}^2\ln n}{n^{c_3}}  \leq
\frac{{K_n}^2}{P_n}
 \leq \frac{{\nu_n}^2}{\ln n} $ both for all $n$
sufficiently large. 

We then derive the constraint on $P_n$. %
%
%
 From condition (\ref{thm:s_Kn}), we obtain
 $c_2 \sqrt{\frac{P_n \ln n}{n^{c_3}}} \leq \nu_n \sqrt{\frac{P_n}{\ln
 n}}$, and $c_2 \sqrt{\frac{P_n \ln n}{n^{c_3}}} \leq {\frac{ c_4
P_n}{n\ln n}}$, both for all $n$ sufficiently large.
 The former constraint leads to $\nu_n \geq \frac{c_2 \ln
n}{n^{c_3/2}}$, which with $\nu_n = o(1)$ can be easily satisfied by
finding suitable $\nu_n$ (e.g., $\nu_n = \frac{c_2 \ln
n}{n^{c_3/3}}$) in view of $c_3
> 0$. It is easy to see that the latter constraint
yields (vi) $P_n \geq {c_2}^2 {c_4}^{-2} n^{2-c_3} (\ln n)^3$, for
all $n$ sufficiently large.


We now present the constraint $K_n$. From condition (vi) and
$\frac{{K_n}^2}{P_n} \geq \frac{{c_2}^2\ln n}{n^{c_3}} $ (derived
from (\ref{thm:s_Kn})), it holds that (vii) $K_n \geq c_2 \sqrt{\frac{P_n \ln n}{n^{c_3}}}  
  \geq
  {c_2}^2 {c_4}^{-1}n^{1-c_3}(\ln n)^2 $ for all $n$
sufficiently large. 
%
%
%


To explain the practicality of (\ref{thm:s_Kn}), it suffices to show
constraints (iv)--(vii) above are all satisfied in practice. As long
as $c_2,c_4
>0$ and $0<c_3<1$ hold,
 constants $c_2,c_3$ and $c_4$ can be specified arbitrarily.
 For $c_3$ close to $1$, we know
that by (vi), the key pool size $P_n$ can be the node number $n$
multiplied by a small fractional power order of $n$; and by (vii),
the key ring size $K_n$ can have
 a small fractional power order of
$n$. These $K_n$ and $P_n$ are practical. In addition, the condition
that $\frac{{K_n}^2}{P_n} \cdot n^{1/3}\ln n$ \emph{either} is
bounded for all $n$ \emph{or} converges to $\infty$ as $n \to
\infty$ is imposed to avoid the degenerate situation where as $n \to
\infty$, the sequence $\frac{{K_n}^2}{P_n} \cdot n^{1/3}\ln n$ does
not approach to $\infty$ yet has a subsequence tending to $\infty$.

In particular, for $P_n = \Theta\big(n^{1+\varepsilon_1}\big)$ and
$K_n = \Theta\big(n^{\varepsilon_2}\big)$ with $\varepsilon_1$ and
$\varepsilon_2$ satisfying $0<\varepsilon_1<1$ and
$\frac{\varepsilon_1}{2} < \varepsilon_2 < \varepsilon_1$, we can
ensure (iv)--(vii) with suitably selected $c_2,c_3$ and $c_4$; i.e.,
(\ref{thm:s_Kn}) holds. Such values of $P_n$ and $K_n$ are very
practical with $\varepsilon_1$ and $\varepsilon_2$ arbitrarily
small. We set $c_3
> 1 + \varepsilon_1 - 2\varepsilon_2$, and specify
 $c_2$ and $c_4$ appropriately (recall that $0<c_3<1$ and $c_2,c_4
>0$ also have to be hold as conditions in Theorem \ref{thm:s}). As
(vi) and (vii) are implied by (\ref{thm:s_Kn}), which is equivalent
to the combination of (iv) and (v), we only need to show (iv) and
(v) as follows. For $P_n = \Theta\big(n^{1+\varepsilon_1}\big)$ and
$K_n = \Theta\big(n^{\varepsilon_2}\big)$, with $\varepsilon_2 <
\varepsilon_1$, then $\frac{K_n}{P_n} =
\Theta\big(n^{-1+\varepsilon_2-\varepsilon_1}\big)$, so (iv) holds
for arbitrary constant $c_4$. Moreover, due to
$2\varepsilon_2>\varepsilon_1$ and $c_3 > 1 + \varepsilon_1 -
2\varepsilon_2$, then $\frac{{K_n}^2}{P_n} = \Theta\big(n^{-(1 +
\varepsilon_1 - 2\varepsilon_2)}\big) $ so $ \frac{{K_n}^2}{P_n}
\geq \frac{{c_2}^2\ln n}{n^{c_3}} $ for all $n$ sufficiently large
with arbitrary constant $c_2$; and because of $1 + \varepsilon_1 -
2\varepsilon_2>1-\varepsilon_1>0$, \h $\frac{{K_n}^2}{P_n}  =
\Theta\big(n^{-(1 + \varepsilon_1 - 2\varepsilon_2)}\big)$, so
$\frac{{K_n}^2}{P_n}
 \leq \frac{{\nu_n}^2}{\ln n} $ for all $n$
sufficiently large after we find suitable $\nu_n = o(1)$; e.g.,
$\nu_n = \Theta\big(n^{-(1 + \varepsilon_1 -
2\varepsilon_2)/3}\big)$. Therefore, we have demonstrated both (iv)
and (v), thus validating (\ref{thm:s_Kn}).%

\section{NUMERICAL EXPERIMENTS} \label{sec:expe}

\begin{figure}[!t]
  \centering
\psfrag{Y}{{{{\fontsize{9.5}{11} \selectfont $r$}}}}
\psfrag{Z}{{\hspace{-11.9pt}${\mathsmaller{\mathbb{P}[G ( n, {K},
{P}, r, \mathcal{A} )\textrm{ is connected.}]}}$}}
 \includegraphics[width=0.4\textwidth]{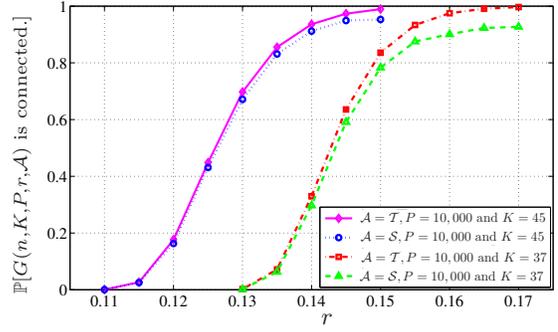}
\caption{A plot of the empirical probability that graph $G ( n, {K},
{P}, r, \mathcal{A} )$ (i.e., $G(n, \theta, \mathcal{A})$) is
connected as a function of $r$ with $n=2,000$, where $\mathcal{A}$
is either the unit torus $\mathcal{T}$ or the unit square
$\mathcal{S}$.\vspace{10pt}} \label{fig}
\end{figure}

We present numerical simulation in the non-asymptotic regime to
support our asymptotic results. We write graph $G(n, \theta_n,
\mathcal{A})$ as $G ( n, {K}_n, {P}_n, r_n, \mathcal{A}
)$. %
%
%
%
  In Figure \ref{fig}, we depict the probability that graph
$G ( n, {K}, {P}, r, \mathcal{A} )$ (i.e., $G(n, \theta,
\mathcal{A})$) is connected, where $\mathcal{A}$ is either the unit
torus $\mathcal{T}$ or the unit square $\mathcal{S}$; and the
subscript $n$ is removed since we fix the number of nodes at
$n=2,000$ in all experiments. For each pair $(\mathcal{A}, K, P,
r)$, we generate $500$ independent samples of $G ( n, {K}, {P}, r,
\mathcal{A} )$ and count the number of times that the obtained
graphs are connected. Then the count divided by $500$ becomes the
empirical probability for connectivity. As illustrated, we observe
the evident threshold behavior in the probability that $G ( n, {K},
{P}, r, \mathcal{A} )$ is connected as such probability transitions
from zero to one as $r$ varies slightly from a certain value.

\section{\secfntx APPLICATION IN FREQUENCY HOPPING}\label{sec:disu:Mobility:fh}


Frequency hopping is a classic approach for transmitting wireless
signals by switching a carrier among different frequency channels.
Frequency hopping offers improved communication resistance to
narrowband interference, jamming attacks,  and signal interception
by eavesdroppers. It also enables more efficient bandwidth
utilization than fixed-frequency transmission
\cite{frequency_hopping1}. For these reasons, military radio
systems, such as \verb|HAVE QUICK| and \verb|SINCGARS|
\cite{havequick},  use frequency hopping extensively. A typical
method of implementing frequency hopping is for the sender and
receiver to first agree on a \emph{secret seed} and a
\emph{pseudorandom number generator} (PRNG). Then the seed is input
to the PRNG by both the sender and the receiver to produce a
sequence of pseudo-random frequencies, each of which is used for
communication in a time interval \cite{frequency_hopping1}.

We consider a wireless network of $n$ nodes where nodes establish
shared secret seeds for frequency hopping as follows. Each node
uniformly and independently selects $K_n$ secret seeds out of a
\textit{secret pool} consisting of $P_n$ secret seeds. Two nodes can
communicate with each other via frequency hopping if and only if
they share at least one secret seed \emph{and}  are within each
other's transmission range. Two nodes can derive a \emph{unique}
seed from the shared seeds in several ways. For example, the unique
seed could be the cryptographic hash of the concatenated seeds
shared between two nodes \cite{adrian}. Alternately, if two
nodes $u$ and $v$ share a 
seed $k_{uv}$ (which might also be shared by other pairs of nodes),
they can establish a probabilistically unique secret seed
$H(u,v,k_{uv})$, where the two node identities are ordered and $H$
is an entropy-preserving cryptographic hash function. 

The above way of bootstrapping seeds has the following advantages.
First, without knowledge of a PRNG seed, an adversary cannot predict
in advance the frequency that two nodes will use. In addition, each
communicating pair of nodes can generate a secret seed that differs
from the seed that another nearby node pair uses. Then it is also
likely that distinct communicating node pairs located in the same
vicinity utilize different frequencies. Thus, without any additional
coordination protocol to avoid using the same frequency, distinct
communicating node pairs nearby could work simultaneously without
causing co-channel interference.

Now we construct a graph $G_f$ based on the above scenario. Each of
the $n$ wireless nodes represents a node in $G_f$. There exists an
edge between two nodes in $G_f$ if and only if they can communicate
with each other via frequency hopping; i.e., they share a secret
seed and are in communication range with each other. Therefore, if
all $n$ nodes are uniformly and independently deployed in a network
area $\mathcal{A}$, which is either a unit torus $\mathcal{T}$ or a
unit square $\mathcal{S}$, and all nodes have the same transmission
range $r_n$, then $G_f$ is exactly $G(n, \theta_n, \mathcal{T})$
when $\mathcal{A}=\mathcal{T}$ and $G(n, \theta_n, \mathcal{S})$
when
$\mathcal{A}=\mathcal{S}$. 
 Our zero--one laws on connectivity of $G(n,
\theta_n, \mathcal{T})$ and $G(n, \theta_n, \mathcal{S})$, allow us
to find the network parameters under which $G_f$ is connected. This
provides useful guideline for the design of large-scale wireless
networks with frequency hopping.

%
%

\section{RELATED WORK} \label{related}


\begin{figure}[t]
 \psfrag{A}{{$\hspace{-20pt}\mathsmaller{{\lim\limits_{n \to \infty} \big({\ln \frac{{K_n}^2}{P_n}}\big/{\ln n}\big)}}$}}
\psfrag{B}{{\hspace{-15pt}$\mathsmaller{{\lim\limits_{n \to \infty}
\big\{ \pi \big[{r_n^{*}(\mathcal{S})}\big]^2 \cdot
\frac{{K_n}^2}{P_n} \big/\big(\frac{\ln n}{n}\big)\big\}}}$}}
\psfrag{C}{\centerhack{$\mathsmaller{{-1}}$}}
\psfrag{D}{\centerhack{$\mathsmaller{{-1/3}}$}}
\psfrag{E}{{$\mathsmaller{{2\pi}}$}}
\psfrag{Y}{\centerhack{$\mathsmaller{{0}}$}}
\psfrag{Z}{\centerhack{$\mathsmaller{{0}}$}}
\psfrag{G}{{$\mathsmaller{{8}}$}} \psfrag{H}{{$\mathsmaller{{4}}$}}
\psfrag{I}{{\hspace{-2pt}$\mathsmaller{{4/3}}$}}
\psfrag{M}{{$\mathsmaller{{1}}$}} \psfrag{K}{{\fontsize{7}{8}
\selectfont \hspace{25pt} Upper bounds of MFCS result
\cite{Krzywdzi}}} \psfrag{N}{{\fontsize{7}{8} \selectfont }}
\psfrag{O}{{\fontsize{7}{8} \selectfont \hspace{25pt} Upper bounds
of ISIT result \cite{ISIT_RKGRGG}}} \psfrag{P}{{\fontsize{7}{8}
\selectfont }} \psfrag{J}{{\fontsize{7}{8} \selectfont Exact
values}} \psfrag{Q}{{\fontsize{7}{8} \selectfont }}
    \hspace{-17pt}\includegraphics[width=.9\columnwidth]{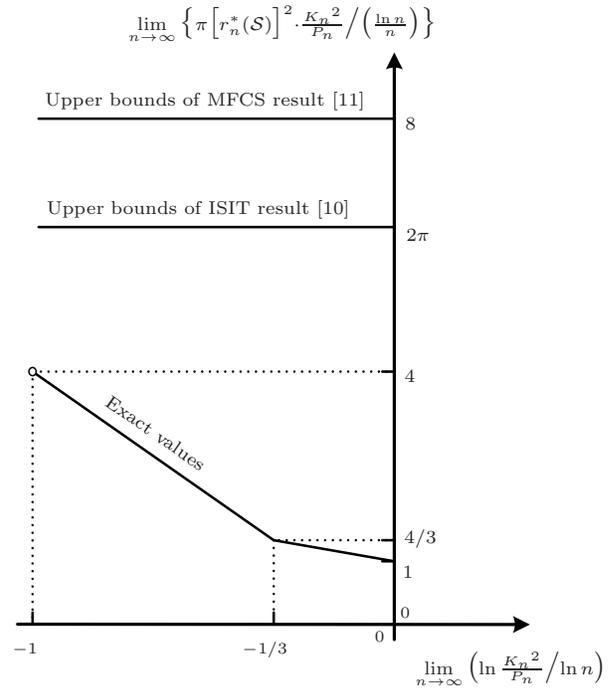}
  \caption{A comparison of the connectivity results for graph
  $G(n, \theta_n, \mathcal{S})$, the intersection of random key graph $G_{RKG}(n, K_n, P_n)$ and
random geometric
 graph $G_{RGG}(n,r_n,\mathcal {A})$, where $r_n^{*}(\mathcal{S})$
is the critical transmission range for connectivity in $G(n,
\theta_n, \mathcal{S})$.\vspace{15pt}} \label{curve}
  \end{figure}

Yi \cite{4151628} \emph{et al.} consider graph $G(n, \theta_n,
\mathcal{A})$, where the network region $\mathcal{A}$ is either a
disk $\mathcal{D}$ or a square $\mathcal{S}$, each of unit
 area. They show that for graph $G(n, \theta_n, \mathcal{D})$ or $G(n, \theta_n, \mathcal{S})$, if $\pi {r_n}^2 \cdot \frac{{K_n}^2}{P_n} = \frac{\ln n +
\alpha}{n}$ and $\frac{{K_n}^2}{P_n} = \omega\big(\frac{1}{\ln
n}\big)$, the number of isolated nodes asymptotically follows a
Poisson distribution with mean $e^{-\alpha}$. Pishro-Nik \emph{et
al.} \cite{Pishro} also obtain such result on asymptotic Poisson
distribution with condition $\frac{{K_n}^2}{P_n} =
\omega\big(\frac{1}{\ln n}\big)$ generalized to $\frac{{K_n}^2}{P_n}
= \Omega\big(\frac{1}{\ln n}\big)$. They further investigate
connectivity in graph $G(n, \theta_n, \mathcal{S})$. In practical
WSNs, $K_n$ is expected to be several orders of magnitude smaller
than $P_n$, so it often holds that $\frac{{K_n}^2}{P_n} =
o\big(\frac{1}{\ln n}\big)$, which is not addressed in the two work
above \cite{4151628,Pishro} and is addressed in our theorems.
Recently, for graph $G(n, \theta_n, \mathcal{S})$, Krzywdzi\'{n}ski
and Rybarczyk \cite{Krzywdzi} and Krishnan \emph{et al.}
\cite{ISIT_RKGRGG} obtain connectivity results, covering the case of
$\frac{{K_n}^2}{P_n} = o\big(\frac{1}{\ln n}\big)$. We elaborate
their theoretical findings below and explain that our results
significantly improve theirs. Krzywdzi\'{n}ski and Rybarczyk
\cite{Krzywdzi} present that in $G(n, \theta_n, \mathcal{S})$ on
$\mathcal{S}$, if $\pi {r_n}^2 \cdot \frac{{K_n}^2}{P_n} \geq
\frac{8\ln n}{n}$ with $K_n \geq 2$ and $P_n = \omega(1)$,\\ then
$G(n, \theta_n, \mathcal{S})$ is almost surely\footnote{An event
occurs \emph{almost surely} if its probability approaches to 1 as
$n\to \infty$.} connected. Krishnan \emph{et al.} \cite{ISIT_RKGRGG}
demonstrate that if $\pi {r_n}^2 \cdot \frac{{K_n}^2}{P_n} \geq
\frac{2\pi \ln n}{n}$ with $K_n = \omega(1)$ and
$\frac{{K_n}^2}{P_n} = o(1)$, then $G(n, \theta_n, \mathcal{S})$ is
almost surely connected. Both only provide upper bounds on
${{\lim\limits_{n \to \infty} \big\{ \pi
\big[{r_n^{*}(\mathcal{S})}\big]^2 \cdot \frac{{K_n}^2}{P_n}
\big/\big(\frac{\ln n}{n}\big)\big\}}}$, with one being $8$ and the
other being $2\pi$, where $r_n^{*}(\mathcal{S})$ is the critical
transmission range for connectivity in $G(n, \theta_n,
\mathcal{S})$. In this paper, we determine the exact value of this
limit by deriving $r_n^{*}(\mathcal{S})$. As illustrated in Figure
\ref{curve}, we plot the term ${{\lim\limits_{n \to \infty} \big\{
\pi \big[{r_n^{*}(\mathcal{S})}\big]^2 \cdot \frac{{K_n}^2}{P_n}
\big/\big(\frac{\ln n}{n}\big)\big\}}}$ with respect to
${{\lim\limits_{n \to \infty} \big({\ln
\frac{{K_n}^2}{P_n}}\big/{\ln n}\big)}}$. The curve of the exact
values is based on our result (\ref{d}) in Section \ref{sec:ctr}.



For random key graph $G_{RKG}(n,K_n,P_n)$, Blackburn and Gerke
\cite{r1}, Rybarczyk \cite{ryb3}, and Ya\u{g}an and Makowski
\cite{yagan} establish zero--one laws for its connectivity. In
particular, Rybarczyk's result is that with $K_n \geq 2$ for all $n$
sufficiently large and $\frac{{K_n}^2}{P_n}=\frac{\ln n +
{\alpha_n}}{n}$, then graph $G_{RKG}(n,K_n,P_n)$ is almost surely
connected (resp., disconnected) if $\lim\limits_{n \to
\infty}{\alpha_n} =\infty$ (resp., $\lim\limits_{n \to
\infty}{\alpha_n} =-\infty$). Rybarczyk \cite{zz} also shows
zero--one laws for $k$-connectivity, where $k$-connectivity means
that the graph remains connected despite the removal of any $(k-1)$
nodes.

Random geometric graph $G_{RGG}(n, r_n, \mathcal{A})$ has been
widely studied due to its application to wireless networks. Gupta
and Kumar \cite{Gupta98criticalpower} show that when $\mathcal{A}$
  is a unit-area disk $\mathcal{D}$ and $\pi {r_n}^2 = \frac{\ln n + \alpha_n}{n}$,
 $G_{RGG}(n, r_n, \mathcal{D})$ is almost surely connected if and only
if $\lim_{n\to\infty}\alpha_n=\infty$.
    Penrose \cite{penrose} explores
    $k$-connectivity in $G_{RGG}(n,r, \mathcal{A})$,
     where $\mathcal{A}$ is a $d$-dimensional unit cube with $d\geq 2$. For $\mathcal{A}$ being
     the unit torus $\mathcal{T}$, he
     obtains that
    with $\rho_n$ denoting the minimum $r_n$ to ensure $k$-connectivity in
     $G_{RGG}(n, r_n, \mathcal{T})$, where $k \geq 1$, then the probability that $\pi {\rho_n}^2$ is
at most $\ln n + (k-1)\ln \ln n - \ln [(k-1)!] + \alpha
 $ asymptotically converges to $e^{- e^{-\alpha}} $. Li \emph{et al.}
 \cite{Li:2003:FTD:778415.778431} prove that
with $k \geq 2$, to have graph $G_{RGG}(n,r, \mathcal{S})$
asymptotically $k$-connected with probability at least $e^{-
e^{-\alpha}} $ for some $\alpha$, a sufficient condition is that the
term $\pi {r_n}^2 $ is at least $\ln n + (2k-3)\ln \ln n - 2\ln
[(k-1)!] + 2\alpha
 $;
and a necessary condition is that $\pi {r_n}^2 $ is no less than
$\ln n + (k-1)\ln \ln n - \ln [(k-1)!] + \alpha$. For $k \geq 2$,
Wan \emph{et al.} \cite{wanAsymptoticCritical} determine the exact
formula of $r_n$ such that graph $G_{RGG}(n,r, \mathcal{S})$ or
$G_{RGG}(n, r_n, \mathcal{D})$ is asymptotically $k$-connected with
 probability $e^{- e^{-\alpha}} $, where as noted above, $\mathcal{D}$ is a disk of unit
 area.

%
%

%
%
%
%
%
%

\section{CONCLUSION}
\label{sec:Conclusion}

We establish the first and {sharp zero--one laws} for connectivity
in WSNs employing the widely-used Eschenauer--Gligor key
pre-distribution scheme under transmission constraints. Such
zero--one laws significantly improve recent results
\cite{ISIT_RKGRGG,Krzywdzi} in the literature. Our theoretical
findings are confirmed via numerical experiments, and are applied to
frequency hopping of wireless networks.

%



\bibliographystyle{abbrv}
\bibliography{related}

\normalsize

\appendix

\section{USEFUL LEMMAS}

\begin{lem}[{Palm's Theory \cite{citeulike:505396}}] \label{lem:palm}

Consider a Poisson process with density $\lambda$ counting the
number of events. If independent of the Poisson process, each event
further does not survive with probability $p$, then the number of
survived events is a Poisson variable with mean $p\lambda$.

\end{lem}

We defer the proofs of Lemmas \ref{lem:IxIxIy}--\ref{lemu} to
Appendix \ref{sec:prf:lemmas}. As detailed in Section
\ref{sec:poisson} later, we demonstrate the zero--law for graph
$G(n, \theta_n, \mathcal{A})$ by proving the same result for its
Poissonized version, graph $G_{\textrm{Poisson}}(n, \theta_n,
\mathcal{A})$, where the only difference between
$G_{\textrm{Poisson}}(n, \theta_n, \mathcal{A})$ and $G(n, \theta_n,
\mathcal{A})$ is that the node distribution of the former is a
homogeneous Poisson point process with intensity $n$ on
$\mathcal{A}$ while that of the latter is a uniform $n$-point
process. Then we present Lemma \ref{lem:IxIxIy} on
$G_{\textrm{Poisson}}(n, \theta_n, \mathcal{A})$.

\begin{lem} \label{lem:IxIxIy}
In graph $G_{\textrm{Poisson}}(n, \theta_n, \mathcal{A})$, let $I_x$
be the event that node $v_x$ is isolated, and $D_{r_n}(\hat{v_x})$
be the intersection of $\mathcal {A}$ and the disk centered at
position $\hat{v_x} \in \mathcal {A}$ with radius $r_n$. We have
\begin{align}
 \mathbb{P}[I_x]  &= \int_{ \mathcal{A} }
 e^{-n p_s |D_{r_n}(\hat{v_x})|} \, \textrm{d}\hat{v_x}; \label{eq_Ix}
\end{align}
and with $\phi_u $ denoting $ {\mathbb{P}}\big[\hspace{1pt} K_{{x}
j} \bcap K_{{y} j} \boldsymbol{\mid} (|S_{xy}| = u)
\hspace{1pt}\big]$, where $u = 0, 1, \ldots, K_n$, then for $u = 1,
2, \ldots, K_n$,
\begin{align}
& \mathbb{P}\big[I_x \bcap I_y \boldsymbol{\mid} (|S_{xy}| = u)\big]
\nonumber \\
&   = \int_{\mathcal {A}} \int_{\mathcal {A}\setminus
D_{r_n}(\hat{v_x})} \nonumber \\
& \quad
 \hspace{-1pt} e^{-n\{ p_s |D_{r_n}(\hat{v_x})| + p_s |D_{r_n}(\hat{v_y})| - \phi_u |D_{r_n}(\hat{v_x}) \hspace{1pt}\cap\hspace{1pt} D_{r_n}(\hat{v_y})|\}}
  \textrm{d}\hat{v_x} \textrm{d}\hat{v_y} , \label{eq_IxIy_Sxyu}
\end{align}
and
\begin{align}
& \mathbb{P}\big[I_x \bcap I_y \boldsymbol{\mid} (|S_{xy}| = 0)\big]
\nonumber \\
&   = \int_{\mathcal {A}} \int_{\mathcal {A}} \nonumber \\
& \quad
 \hspace{-1pt} e^{-n\{ p_s |D_{r_n}(\hat{v_x})| + p_s |D_{r_n}(\hat{v_y})| - \phi_{\mathlarger{_0}} |D_{r_n}(\hat{v_x}) \hspace{1pt}\cap\hspace{1pt} D_{r_n}(\hat{v_y})|\}}
  \textrm{d}\hat{v_x} \textrm{d}\hat{v_y},\label{eq_IxIy_Sxy0}
\end{align}
with $\phi_{\mathlarger{_0}}$ meaning $\phi_u$ when $u=0$.
\end{lem}

%

\begin{lem} \label{lemt}


Under (\ref{thm:t_Kn}) and (\ref{thm:t:rnKnPn}), for all $n$
sufficiently large, with $ \alpha_n <0$, we obtain $\pi {r_n}^{2}  n
\cdot \frac{{K_n}^2}{P_n} \leq \ln n$, $r_n   = o(1)$, and $\pi
{r_n}^2 {p_s} n  = \ln n +
 \alpha_{n} - O(1)$.

\end{lem}

\begin{lem} \label{lems}


Under (\ref{thm:s_Kn}) and (\ref{thm:s:rnKnPn}), we have the
following: $\ln \frac{n}{p_s} = \Theta(\ln n)$; and for $ |\alpha_n|
= O(\ln \ln n)$, then (a) $r_n  = o(1)$, $\pi {r_n}^2
\cdot\frac{{K_n}^2}{P_n} = \Theta\big(\frac{\ln n}{n}\big)$ and $\pi
{r_n}^2 \cdot p_s = \Theta\big(\frac{\ln n}{n}\big)$; and (b) with
$\delta_n$ for all $n  $ be defined via
\begin{align}
&\pi {r_n}^2 {p_s}\hspace{-1pt}=\hspace{-1pt}\begin{cases}
\hspace{-4pt}\textrm{\fontsize{13}{15} \selectfont $\frac{\ln
\frac{n}{{p_s}} \hspace{2pt}-\hspace{2pt} \ln \ln \frac{n}{{p_s}}
\hspace{2pt}+\hspace{2pt} \delta_{n} }{n}$} ,
&\hspace{-7pt}\textrm{for~} p_s = \omega
\textrm{\large \selectfont $\left( \frac{1}{n^{1/3}\ln n} \right)$}, \\
\vspace{-7pt} \\
 \hspace{-4pt}\textrm{\fontsize{13}{15} \selectfont $\frac{ 4\ln \frac{1}{p_s} \hspace{2pt}
 -\hspace{2pt} 4\ln \ln
\frac{1}{p_s} \hspace{2pt}+\hspace{2pt} \delta_{n}  }{n}$} ,
&\hspace{-7pt}\textrm{for~} p_s  = O \textrm{\large \selectfont
$\left( \frac{1}{n^{1/3}\ln n} \right)$},\end{cases} \label{deltan}
\end{align}
\h $\delta_{n}   =  \alpha_{n} \pm O(1) $.

\end{lem}

%
%


\begin{lem} \label{lemu}

 If $P_n \geq 3K_n $, then 
  for any three distinct nodes $v_x, v_y$ and
$v_z$ and for any $u = 0, 1, \ldots, K_n$,
\begin{align}
\mathbb{P}[({K}_{xz} \cap {K}_{yz} \boldsymbol{\mid} (|S_{xy}| = u)]
& \leq
 \frac{ u{K_n}}{P_n} + \frac{2{K_n}^4}{{P_n}^2}.
\nonumber
\end{align}

\end{lem}

\begin{lem} \label{Poissonization1}
In graph $G_{\textrm{Poisson}}(n, \theta_n, \mathcal{T})$ under
conditions (\ref{thm:t_Kn}) and (\ref{thm:t:rnKnPn}) with
$|\alpha_n| = O(\ln \ln n)$, then
\begin{align}
  & n\mathbb{P}[I_x] = o(n^{\epsilon})\textrm{ for any constant
  }\epsilon>0. \label{epsilon_po_t}
 \end{align}
\end{lem}

\begin{lem} \label{Poissonization2}
In graph $G_{\textrm{Poisson}}(n, \theta_n, \mathcal{S})$ under
conditions (\ref{thm:s_Kn}) and (\ref{thm:s:rnKnPn}) with
$|\alpha_n| = O(\ln \ln n)$, then
\begin{align}
  & n\mathbb{P}[I_x] = o(n^{\epsilon})\textrm{ for any constant
  }\epsilon>0.\label{epsilon_po_s}
 \end{align}
\end{lem}

\begin{lem} \label{dePoissonization}
For $\mathcal{A}$ being $\mathcal{T}$ under conditions
(\ref{thm:t_Kn}) and (\ref{thm:t:rnKnPn}) with $|\alpha_n| = O(\ln
\ln n)$, or $\mathcal{A}$ being $\mathcal{S}$ under conditions
(\ref{thm:s_Kn}) and (\ref{thm:s:rnKnPn}) also with $|\alpha_n| =
O(\ln \ln n)$, then with $m$ denoting $\big\lceil
n-n^{\frac{1}{2}+c_0}\big\rceil$, where $c_0$ is an arbitrary
constant with $0 < c_0 < \frac{1}{2}$, \h
\begin{align}
  & \boldsymbol{\big|} \mathbb{P}\left[\hspace{2pt} G
  (n, \theta_n, \mathcal{A})
  \textrm{ has no isolated node.} \hspace{2pt}\right]
  \nonumber  \\ & - \mathbb{P}\left[\hspace{2pt} G_{\textrm{Poisson}}
  (m, \theta_n, \mathcal{A})
  \textrm{ has no isolated node.} \hspace{2pt}\right] \boldsymbol{\big|} = o(1). \nonumber
\end{align}
\end{lem}

\begin{lem} \label{erg_rgg_conn:t}
 Consider graph
$G_{\hspace{-.5pt}ER\hspace{-.5pt}}(n,\hspace{-.5pt}p_n)\hspace{-1pt}
\bcap \hspace{-1pt} G_{\hspace{-.5pt}RGG\hspace{-.5pt}}(n,
\hspace{-.5pt}r_n, \hspace{-.5pt}\mathcal{T})$,\\ where
$G_{ER}(n,p_n)$ is an Erd\H{o}s--R\'{e}nyi graph; and \\$G_{RGG}(n,
r_n, \mathcal{T})$ is a random geometric graph on a\vspace{1pt} unit
torus $\mathcal{T}$. 
 Let the sequence $\nu_n$\vspace{1pt} for all $n$ be defined through
\begin{align}
 \pi {r_n}^2 p_n & = \frac{\ln n + \nu_{n}}{n}.\nonumber
\end{align}
Then as $n \to \infty$,
\begin{align}
  \mathbb{P}\left[\hspace{-1pt}
  \begin{array}{l} G_{ER}(n,p_n) \\ \bcap G_{RGG}(n, r_n, \mathcal{T}\hspace{1pt})
  \\  \textrm{is connected.}
\end{array} \hspace{-1pt}\right] & \to
\begin{cases} 0, ~~\textrm{if $\lim\limits_{n \to \infty}{\nu_n}
=-\infty$}, \\  1, ~~\textrm{if $\lim\limits_{n \to \infty}{\nu_n}
=\infty$.}\end{cases} \nonumber
\end{align}
\end{lem}

\begin{lem} \label{erg_rgg_conn} \textbf{(\hspace{-0.3pt}\cite[Theorem 2.5 and Proposition 8.5]{Penrose2013}).}
Consider graph $G_{ER}(n,p_n)\bcap G_{RGG}(n, r_n,
\mathcal{S}\hspace{1pt})$ \vspace{1pt}with \\$p_n = o \left(
\frac{1}{\ln n} \right)$, where $G_{ER}(n,p_n)$ is an\vspace{1.5pt}
Erd\H{o}s--R\'{e}nyi graph; and $G_{RGG}(n, r_n, \mathcal{S})$ is a
random geometric graph on a unit square $\mathcal{S}$. With $p_n
\cdot n^{1/3}\ln n$ \emph{either} being bounded for all $n$
\emph{or} converging to $\infty$ as $n \to \infty$, let the sequence
$\nu_n$ for all $n $ be defined through
\begin{align}
\hspace{-117pt} \pi {r_n}^2 p_n & \nonumber
\end{align}\vspace{-7pt}
\small \selectfont
\begin{subnumcases} {=} \hspace{-4pt}\frac{\ln \frac{n}{p_n} - \ln \ln \frac{n}{p_n}
+ \nu_{n}}{n} , \textrm{\normalsize \selectfont~for~\small
\selectfont} p_n = \omega \left(
\frac{1}{n^{1/3}\ln n} \right),   \nonumber  \\
 \hspace{-4pt}\frac{ 4\ln \frac{1}{p_n} - 4\ln \ln
\frac{1}{p_n}  + \nu_{n}}{n}, \textrm{\normalsize
\selectfont~for~\small \selectfont} \begin{array}{l} p_n = O\left(
\frac{1}{n^{1/3}\ln n} \right),\vspace{2pt}
 \\  \textrm{and } r_n = n^{-\Omega(1)}.
\end{array}\nonumber
 \end{subnumcases}
\normalsize \selectfont \vspace{-10pt}
\begin{align}
\label{lem:erg_rgg_rnpn}\vspace{-15pt}
\end{align}
Then as $n \to \infty$,
\begin{align}
  \mathbb{P}\left[\hspace{-1pt}
  \begin{array}{l} G_{ER}(n,p_n) \\ \bcap G_{RGG}(n, r_n, \mathcal{S}\hspace{1pt})
  \\  \textrm{is connected.}
\end{array} \hspace{-1pt}\right] & \to
\begin{cases} 0, ~~\textrm{if $\lim\limits_{n \to \infty}{\nu_n}
=-\infty$}, \\  1, ~~\textrm{if $\lim\limits_{n \to \infty}{\nu_n}
=\infty$.}\end{cases} \nonumber
\end{align}
\end{lem}

\begin{lem} \label{cp_rig_er}

If $K_n = \Omega\left((\ln n)^3\right)$, $\frac{K_n}{P_n} =
O\left(\frac{1}{n \ln n}\right) $ and $\frac{{K_n}^2}{P_n} =
O\left(\frac{1}{ \ln n}\right)$, then there exists $p_n$ with
\begin{align}
p_n & = \frac{{K_n}^2}{P_n} \cdot \left[1 -
 O \left(\frac{1}{ \ln n}\right)\right] \label{eq_pn}
\end{align}
such that for any topology $\mathcal{A}$ and any monotone increasing
graph property\footnote{A graph property is called monotone
increasing if it holds under the addition of edges in a graph.}
$\mathscr{P}$,
\begin{align}
& \mathbb{P}[ \hspace{2pt}G_{RKG}(n, K_n, P_n) \bcap G_{RGG}(n, r_n,
\mathcal{A}) \textrm{ has }\mathscr{P}.\hspace{2pt}] \nonumber
\\ & \quad \geq
 \mathbb{P}[ \hspace{2pt}G_{ER}(n, p_n) \bcap G_{RGG}(n, r_n,
\mathcal{A}) \textrm{ has }\mathscr{P}.\hspace{2pt}] - o(1).
\nonumber
 \end{align}

\end{lem}

\section{ESTABLISHING THE ZERO--LAWS}


We first explain the basic ideas of the proofs.

\subsection{Basic Ideas of the Proofs}

%

\subsubsection{Poissonization and de-Poissonization} \label{sec:poisson}

We demonstrate the zero--laws using the standard Poissonization
technique \cite{citeulike:505396,penrose}. The idea is that the
zero--law for graph $G(n, \theta_n, \mathcal{A})$ follows once we
establish the result with Poissonization; i.e., once we obtain the
zero--law for graph $G_{\textrm{Poisson}}(n, \theta_n,
\mathcal{A})$. See Lemma \ref{dePoissonization} for the rigorous
argument.


\subsubsection{Method of the moments}

We reuse the notation in Lemma \ref{lem:IxIxIy}; i.e., here in graph
$G_{\textrm{Poisson}}(n, \theta_n, \mathcal{A})$, where $\mathcal
{A}$ is the unit torus $\mathcal {T}$ or the unit square $\mathcal
{S}$, let $I_x$ be the event that node $v_x$ is isolated, and
$D_{r_n}(\hat{v_x})$ be the intersection of $\mathcal {A}$ and the
disk centered at position
$\hat{v_x} \in \mathcal {A}$ with radius $r_n$. 

We use the method of the moments for the proof. Note that $n$ is the
expected number of nodes in graph $G_{\textrm{Poisson}}(n, \theta_n,
\mathcal{A})$. By \cite[Fact 1 and Lemma 1]{ZhaoYaganGligor}, the
zero--law is proved once we demonstrate
\begin{align}
\lim\limits_{n \to \infty}n \mathbb{P}[I_x] & = \infty, \label{nPIx}
\end{align}
and
\begin{align}
\mathbb{P}[I_x \bcap I_y] & \leq \big\{\mathbb{P}[I_x]\big\}^2 \cdot
[1+o(1)].  \label{PIxIy}
\end{align}
Below we prove (\ref{nPIx}) and (\ref{PIxIy}), respectively. Note
that given condition $\lim\limits_{n \to \infty}{\alpha_n} =-\infty$
in the zero--laws, we obtain ${\alpha_n} <0$ for all $n$
sufficiently large.

\subsection{Proving the Zero--Law of Theorem \ref{thm:t}} \label{sec:thm:t:zero}


As just noted, we have $\alpha_n < 0$ for all $n$ sufficiently large
so we can use results from Lemma \ref{lemt}.

\subsubsection{Establishing (\ref{nPIx}) on the unit torus $\mathcal {T}$}
\label{sec:t:0:p1}

By Lemma \ref{lem:IxIxIy}, \h
\begin{align}
 \mathbb{P}[I_x]  &= \int_{ \mathcal{T} }
 e^{-n p_s |D_{{r_n}}(\hat{v_x})|} \, \textrm{d}\hat{v_x}. \nonumber
\end{align}
Since $\mathcal {T}$ is a unit torus, \h $D_{{r_n}}(\hat{v_x}) = \pi
{r_n}^2$ for any $\hat{v_x} \in \mathcal {T}$. Then
\begin{align}
 \mathbb{P}[I_x] & = e^{-\pi {r_n}^2 p_s n} \cdot |\mathcal{T}| =
e^{-\pi {r_n}^2
 p_s n}. \label{prIxe}
\end{align}
Using $\pi {r_n}^2 {p_s} n  = \ln n +
 \alpha_{n} - O(1)$ from
Lemma \ref{lemt} in (\ref{prIxe}) and considering $\lim\limits_{n
\to \infty}{\alpha_n} =-\infty$, we obtain
\begin{align}
 n\mathbb{P}[I_x] & = n e^{-\ln n - \alpha_n - O(1)} \to \infty
\textrm{ as }n \to \infty. \nonumber
\end{align}

\subsubsection{Establishing (\ref{PIxIy}) on the unit torus $\mathcal {T}$}

By the law of total probability, it is clear that
\begin{align}
\mathbb{P}[I_x \bcap I_y]  &
  = \sum_{u=0}^{K_n} \mathbb{P}\big[\hspace{1pt}I_x
\bcap I_y \boldsymbol{\mid} (|S_{xy}| = u)\hspace{1pt}\big]
\mathbb{P}[\hspace{1pt}|S_{xy}| = u\hspace{1pt}]. \label{proIxIy}
\end{align}
Applying Lemma \ref{lem:IxIxIy} to (\ref{proIxIy}), we derive
\begin{align}
& \mathbb{P}\big[\hspace{1pt}I_x \bcap I_y \boldsymbol{\mid}
(|S_{xy}| = u)\hspace{1pt}\big]  & \nonumber \\ &
  \leq  \int_{\mathcal {A}} \int_{\mathcal {A}} \nonumber \\
&  ~~e^{-n\{p_s |D_{{r_n}}(\hat{v_x})| + p_s |D_{{r_n}}(\hat{v_y})|
- \phi_u |D_{{r_n}}(\hat{v_x}) \hspace{1pt}\cap\hspace{1pt}
D_{{r_n}}(\hat{v_y})|\}}
 \textrm{d}\hat{v_x} \textrm{d}\hat{v_y}. \label{leq_IxIy}
\end{align}
Here we consider $\mathcal {A}$ as the torus $\mathcal {T}$. For any
$\hat{v_x} \in \mathcal {T}$ and any $\hat{v_y} \in \mathcal {T}$,
we have $D_{{r_n}}(\hat{v_x}) = \pi {r_n}^2$ and
$D_{{r_n}}(\hat{v_y}) = \pi {r_n}^2$. If $\hat{v_y} \in \mathcal
{A}\setminus D_{2{r_n}}(\hat{v_x})$ (i.e., $\hat{v_x}$ and
$\hat{v_y}$ have a distance greater than $2{r_n}$), where
$D_{2{r_n}}(\hat{v_x})$ is the intersection of $\mathcal {T}$ and
the disk centered at $\hat{v_x}$ with radius $2{r_n}$, then
$|D_{r_n}(\hat{v_x}) \bcap D_{r_n}(\hat{v_y})| = 0$; and if
$\hat{v_y} \in D_{2{r_n}}(\hat{v_x})$, then $|D_{r_n}(\hat{v_x})
\bcap D_{r_n}(\hat{v_y})| \leq \pi {r_n}^{2}$. Therefore, from
(\ref{leq_IxIy}),
\begin{align}
  & \mathbb{P}\big[\hspace{1pt}I_x
\bcap I_y \boldsymbol{\mid} (|S_{xy}| = u)\hspace{1pt}\big]
\nonumber \\   &
  \leq 
\mathlarger{\big(}1-4\pi {{r_n}}^2 + 4\pi {r_n}^2 e^{\pi
{r_n}^{2}\phi_u n}\mathlarger{\big)} e^{-2 \pi {r_n}^{2} p_s n}.
\label{ev_pxy}
\end{align}
Substituting (\ref{ev_pxy}) into (\ref{proIxIy}), we obtain
\begin{align}
 &\hspace{-2pt} \mathbb{P}[I_x \bcap I_y] \nonumber \\   & \hspace{-2pt} \leq (1-4\pi {r_n}^2) e^{-2 \pi {r_n}^{2} p_s
n}  \nonumber \\   & \hspace{2pt} + 4\pi {r_n}^2 e^{-2 \pi {r_n}^{2}
p_s n} \sum_{u=0}^{K_n} \Big\{ \mathbb{P}[\hspace{1pt}|S_{xy}| =
u\hspace{1pt}] e^{ \pi {r_n}^{2} \phi_u n} \Big\}.
\label{mathbbIxIy}
\end{align}

Applying ${r_n}  = o(1)$ from Lemma \ref{lemt} to
(\ref{mathbbIxIy}), then (\ref{PIxIy}) is proved once we show
\begin{align}
 \sum_{u=0}^{K_n} \Big\{ \mathbb{P}[\hspace{1pt}|S_{xy}| =
u\hspace{1pt}] e^{ \pi {r_n}^{2} \phi_u n} \Big\}  & = O(1).
\label{sumu0K}
\end{align}
By \cite[Lemma 10]{ZhaoYaganGligor}, $\mathbb{P}[|S_{xy}|=u ]  \leq
\frac{1}{u!} \big(\frac{{K_n}^2}{P_n-K_n}\big)^u$ holds, 
which along with Lemma \ref{lemu} gives rise to
\begin{align}
& \sum_{u=0}^{K_n} \Big\{ \mathbb{P}[\hspace{1pt}|S_{xy}| =
u\hspace{1pt}] e^{
\pi {r_n}^{2} \phi_u n} \Big\}
 \nonumber \\%
& \quad \leq e^{2\pi {r_n}^{2}n \cdot \frac{{K_n}^4}{{P_n}^2} +
\frac{{K_n}^2}{P_n-K_n} \cdot e^{ \pi {r_n}^{2}n\frac{K_n}{P_n}}}.
\label{e2pir}
\end{align}
Given $\frac{{K_n}^2}{P_n}
 \leq \frac{{c_1}^2}{\ln n}$,
 we have $\frac{{K_n}^2}{P_n} = O\big(\frac{1}{\ln n}\big)$
  and thus $\frac{{K_n}^2}{P_n-K_n}  \leq  \frac{2{K_n}^2}{P_n}$ for all $n$ sufficiently
  large. %
%
 In view of $\pi {r_n}^{2}  n \cdot \frac{{K_n}^2}{P_n} \leq \ln n$
from Lemma \ref{lemt} and $K_n\geq \frac{\ln n}{\ln \ln n}$ by
condition (\ref{thm:t_Kn}), it holds that for all $n$ sufficiently
large,
\begin{align}
e^{\pi {r_n}^{2}n\frac{K_n}{P_n}} &   \leq e^{ {K_n}^{-1} \ln n}
 \leq e^{\ln\ln n} = \ln n.  \label{epilnn}
\end{align}
Using (\ref{epilnn}), $\frac{{K_n}^2}{P_n-K_n}  \leq
\frac{2{K_n}^2}{P_n}$, $\pi {r_n}^{2}  n \cdot \frac{{K_n}^2}{P_n}
\hspace{-.7pt}\leq \hspace{-.7pt} \ln n$ and $\frac{{K_n}^2}{P_n}
\hspace{-.7pt}= \hspace{-.7pt}O\big(\frac{1}{\ln n}\big)$ in
(\ref{e2pir}),
we establish (\ref{sumu0K}). 
 As explained before, the proof of
(\ref{PIxIy}) is now completed.

\subsection{Proving the Zero--Law of Theorem \ref{thm:s}}

We will explain that $|\alpha_n|$ can be confined as $ O(\ln \ln
n)$. To see this for the zero--law, it suffices to show
\begin{align}
&\textrm{The \hspace{-.5pt}zero--law\hspace{-.5pt} of\hspace{-.5pt}
Theorem\hspace{-.5pt} \ref{thm:s}\hspace{-.5pt} under
\hspace{-.5pt}$|\alpha_n| = O(\ln \ln n)$}
\Rightarrow \nonumber \\
&  \textrm{The \hspace{-.5pt}zero--law\hspace{-.5pt}
of\hspace{-.5pt} Theorem\hspace{-.5pt} \ref{thm:s}\hspace{-.5pt}
regardless\hspace{-.5pt} of \hspace{-.5pt}} |\alpha_n|
\hspace{-2pt}= \hspace{-2pt}O(\ln \ln n)  . \nonumber
\end{align}

Letting $\widetilde{\alpha}_n $ be $\max\{\alpha_n, - \ln \ln n\} $,
we define $\widetilde{r}_n $ through
\begin{align}
\hspace{-117pt} \pi {\widetilde{r}_n}^2 \cdot
\textrm{\fontsize{9.5}{11} \selectfont $\frac{{K_n}^2}{P_n}$} &
\nonumber
\end{align}\vspace{-10pt}
\begin{align}
&\hspace{-4pt}=\begin{cases} \hspace{-4pt}\textrm{\fontsize{13}{15}
\selectfont $\frac{\ln \frac{n P_{n}}{{K_{n}}^2}
\hspace{2pt}-\hspace{2pt} \ln \ln \frac{n P_{n}}{{K_{n}}^2}
\hspace{2pt}+\hspace{2pt} \widetilde{\alpha}_n }{n}$} ,
&\hspace{-6pt}\textrm{for~} \textrm{\large \selectfont
$\frac{{K_n}^2}{P_n}$} = \omega
\textrm{\large \selectfont $\left( \frac{1}{n^{1/3}\ln n} \right)$}, \\
\vspace{-7pt} \\
 \hspace{-4pt}\textrm{\fontsize{13}{15} \selectfont $\frac{ 4\ln \frac{P_n}{{K_n}^2} \hspace{2pt}
 -\hspace{2pt} 4\ln \ln \frac{P_n}{{K_n}^2} \hspace{2pt}+\hspace{2pt}
\widetilde{\alpha}_n  }{n}$} , &\hspace{-6pt}\textrm{for~}
\textrm{\large \selectfont $\frac{{K_{n}}^2}{P_{n}}$}  = O
\textrm{\large \selectfont $\left( \frac{1}{n^{1/3}\ln n}
\right)$}.\end{cases} \label{widetld}
\end{align}
It is clear that $r_n \leq \widetilde{r}_n$. We write graph $G(n,
\theta_n, \mathcal{S})$ as $G ( n, {K}_n, {P}_n, r_n, \mathcal{S}
)$. Then we can construct graph $G ( n, {K}_n, {P}_n,
\widetilde{r}_n, \mathcal{S} )$
 as follows such that it is a
supergraph of $G ( n, {K}_n, {P}_n, r_n, \mathcal{S} )$. In $G ( n,
{K}_n, {P}_n, r_n, \mathcal{S} )$, with each node increasing its
transmission range from $r_n$ to $\widetilde{r}_n$, then the graph
becomes $G ( n, {K}_n, {P}_n, \widetilde{r}_n, \mathcal{S} )$.

For the zero--law, we consider $\lim\limits_{n \to \infty}{\alpha_n}
=-\infty$, which yields $\lim\limits_{n \to
\infty}{\widetilde{\alpha}_n} =-\infty$ and $|\widetilde{\alpha}_n|
= O(\ln \ln n)$. If we have the zero--law of Theorem \ref{thm:s}
under $|\alpha_n| = O(\ln \ln n)$, then even $|\alpha_n| = O(\ln \ln
n)$ does not hold, in view of $\lim\limits_{n \to
\infty}{\widetilde{\alpha}_n} =-\infty$ and $|\widetilde{\alpha}_n|
= O(\ln \ln n)$, we apply the zero--law to graph $G ( n, {K}_n,
{P}_n, \widetilde{r}_n, \mathcal{S} )$ and obtain that under
(\ref{thm:s_Kn}) and (\ref{widetld}), graph $G ( n, {K}_n, {P}_n,
\widetilde{r}_n, \mathcal{S} )$ is disconnected almost surely. Then
as a subgraph of graph $G ( n, {K}_n, {P}_n, \widetilde{r}_n,
\mathcal{S} )$, graph $G ( n, {K}_n, {P}_n, r_n, \mathcal{S} )$ is
also disconnected. Hence, we obtain the zero--law of Theorem
\ref{thm:s} regardless of the condition $|\alpha_n| =  O(\ln \ln
n)$.



%
%

%
%

  \begin{figure}[!t]
\begin{center}
  \includegraphics[scale=.8]{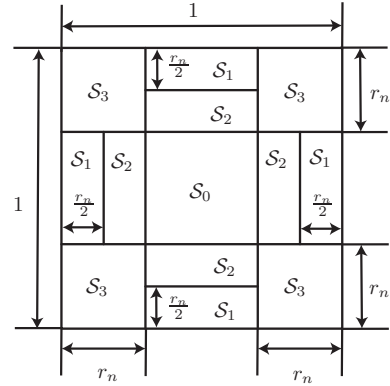}
  \caption{We partition the unit square $\mathcal{S}$ into
  $\mathcal{S}_0, \mathcal{S}_1, \mathcal{S}_2$
 and $\mathcal{S}_3$. Note that each of $\mathcal{S}_1, \mathcal{S}_2$
 and $\mathcal{S}_3$ has four parts. \vspace{5pt}}\label{partition}
  \end{center}
  \end{figure}

\subsubsection{Establishing (\ref{nPIx}) on the unit square $\mathcal
{S}$} \label{sec:sq}

By Lemma \ref{lem:IxIxIy}, $\mathbb{P}[I_x] = \int_{ \mathcal{S} }
 e^{-n p_s D_{{r_n}}(\hat{v_x})} \, \textrm{d}\hat{v_x}$ holds.
 To compute $\mathbb{P}[I_x]$ based on this, we partition $\mathcal{S}$ in a
 way similar to that by Li \emph{et al.} \cite{Li:2003:FTD:778415.778431} and Wan \emph{et al.} \cite{wanAsymptoticCritical}.
Specifically, $\mathcal{S}$ is divided into $\mathcal{S}_0,
\mathcal{S}_1, \mathcal{S}_2$
 and $\mathcal{S}_3$, respectively, as illustrated in Figure
 \ref{partition} (note
that $r_n < \frac{1}{2}$ \n due to $r_n = o(1)$ by Lemma
\ref{lems}). $\mathcal{S}_0$ consists of all points each with a
distance greater than ${r_n}$ to its nearest edge of $\mathcal{S}$,
whereas $\mathcal{S}_3$ is the area in which each point has
distances no greater than ${r_n}$ to at least two edges of
$\mathcal{S}$. We further divide $ \mathcal{S} \setminus
\{\mathcal{S}_0 \cup \mathcal{S}_3\}$ into $\mathcal{S}_1$ and
$\mathcal{S}_2$ as follows. In $ \mathcal{S} \setminus
\{\mathcal{S}_0 \cup \mathcal{S}_3\}$, $\mathcal{S}_1$ compromise
points whose distance to the nearest edge of $\mathcal{S}$ is no
greater than $\frac{{r_n}}{2}$, while the remaining area is
$\mathcal{S}_2$; i.e., $\mathcal{S}_2 = \mathcal{S} \setminus
\{\mathcal{S}_0 \cup \mathcal{S}_1 \cup \mathcal{S}_3\}$.

For $i = 0, 1, 2, 3$, we define 
\begin{align}
 T_i:  &=  \int_{ \mathcal{S}_i }
 e^{-n p_s |D_{{r_n}}(\hat{v_x})|} \, \textrm{d}\hat{v_x}.
 \label{defTi}
\end{align}
 Then it is clear that $\mathbb{P}[I_x] =\sum_{i=0}^{3}T_i$.
 From (\ref{deltan}) in Lemma \ref{lems}, there exists
 $\delta_{n}$ with $\delta_{n} =  \alpha_{n} \pm O(1)$ such that (i) if $p_s =  \omega \left(
\frac{1}{n^{1/3}\ln n} \right)$, then
\begin{align}
\pi {r_n}^2 {p_s} & = \frac{\ln \frac{n}{{p_s}} - \ln \ln
\frac{n}{{p_s}} + \delta_{n} }{n}; \label{casei}
 \end{align}
and (ii) if $p_s  =  O \left( \frac{1}{n^{1/3}\ln n} \right)$, then
\begin{align}
\pi {r_n}^2 {p_s} & = \frac{ 4\ln \frac{1}{p_s} - 4\ln \ln
\frac{1}{p_s} + \delta_{n} }{n}. \label{caseiirp}
 \end{align}
We explain below in detail that in case (i), $\lim\limits_{n \to
\infty}T_1 = \infty$ follows, yielding $\lim\limits_{n \to
\infty}\mathbb{P}[I_x] = \infty$ (i.e., (\ref{nPIx})). 

To evaluate $T_1$, we introduce some notation as follows. For any
position
 $ \hat{v_x} \in \mathcal{S}_1 $, we let
the distance from $\hat{v_x}$ to the nearest edge of the square
$\mathcal{S}$ be $g$, where $0 \leq g \leq \frac{{r_n}}{2}$. For $
\hat{v_x} \in \mathcal{S}_1$, clearly $|D_{{r_n}}(\hat{v_x})| $ is
determined given $g$; and we denote it by $H(g)$. As used before, we
have Lagrange's notation for differentiation; namely, the first and
second derivatives of a function $f$ are denoted by $f'$ and $f''$,
respectively. It is easy to derive
\begin{align}
H(g) &=  [\pi - \arccos ({g}/{{r_n}})]{r_n}^2 + g\sqrt{{r_n}^2 -
g^2},  \label{h} \\ H'(g) & = 2\sqrt{{r_n}^2 - g^2}, \label{h2}
 \end{align}
 and
 \begin{align}
H''(g) &= - 2{g}/{\sqrt{{r_n}^2 - g^2}}. \label{h3}
 \end{align}


 Since $\mathcal{S}_1$ consists of four
rectangles, each of which has length $1-2{r_n}$ and width
$\frac{{r_n}}{2}$, \f
\begin{align}
T_1  \hspace{-1.5pt}&=\hspace{-1.5pt} \int_{ \mathcal{S}_1}
\hspace{-1.5pt} e^{-p_s n |D_{{r_n}}(\hat{v_x})|} \,\textrm{d}
\hat{v_x} \hspace{-1.5pt}=\hspace{-1.5pt} 4
(1\hspace{-1.5pt}-\hspace{-1.5pt}2{r_n}) \int_{ 0}^{
\frac{{r_n}}{2}} \hspace{-1.5pt} e^{-p_s n H(g)} \, \textrm{d}g.
   \label{go8}
 \end{align}
For simplicity, we write $H(g)$ as $H$. Then
\begin{align}
 &
 e^{-p_s n H } \, \textrm{d}g \nonumber \\ & =
  -( p_s  n)^{-1} (H')^{-1} \, \textrm{d} e^{-p_s n H }
\nonumber \\ &  = - ( p_s  n )^{-1}
 \big\{\textrm{d}\big[(H')^{-1}e^{-p_s n H }
 \big]\hspace{-1pt}-\hspace{-1pt}e^{-p_s n H } \, \textrm{d}(H')^{-1} \big\}
   \nonumber \\ &  = ( p_s  n )^{-1}
 \big\{\textrm{d}\big[\hspace{-2pt}-\hspace{-2pt}(H')^{-1}e^{-p_s n H }
 \big]\hspace{-1pt}-\hspace{-1pt}(H')^{-2}H'' e^{-p_s n H }
 \textrm{d}g\big\}.
   \label{go8t1}
 \end{align}
From (\ref{go8t1}) and $H'' \leq 0$,
\begin{align}
&   \int_{ 0}^{ \frac{{r_n}}{2}}
 e^{-p_s n H(g)} \, \textrm{d}g   \geq \frac{ e^{-p_s n H(0) }}{p_s n H'(0)}
   - \frac{ e^{-p_s n H\left(\frac{{r_n}}{2}\right) }}
   {p_s n H'\left(\frac{{r_n}}{2}\right)} . \label{it3mbnwup}
 \end{align}

From (\ref{h}) and (\ref{h2}), then $H(0)  = \pi {r_n}^2 / 2$,
$H'(0) = 2 r_n$, $H\left(\frac{{r_n}}{2}\right) =
\big(\frac{2}{3}\pi  + \frac{\sqrt{3}}{4}\big)  {r_n}^2 $ and
$H'\left(\frac{{r_n}}{2}\right) = \sqrt{3} r_n$. Using these and
$\pi {r_n}^2 {p_s} n  = \Theta(\ln n)$ from Lemma \ref{lems} in
(\ref{it3mbnwup}), 
 we derive
%
%
\begin{align}
  \int_{ 0}^{ \frac{{r_n}}{2}}
 e^{-p_s n H(g)} \, \textrm{d}g & \geq
 \frac{ e^{-\pi {r_n}^2 p_s n / 2}}{2 r_n p_s n} \cdot [1-o(1)], \nonumber
 \end{align}
which along with $r_n = o(1)$ from Lemma \ref{lems} is applied to
(\ref{go8}) so that 
%
%
  \begin{align}
 \mathbb{P}[I_x]  \geq  T_1  & \geq
 2 (r_n p_s n)^{-1} e^{-\pi {r_n}^2 p_s n / 2} \cdot [1-o(1)].\label{pixlo}
  \end{align}
%

From (\ref{casei}), we get
\begin{align}
e^{-\pi {r_n}^2 p_s n} & = e^{-\ln \frac{n}{p_s} + \ln \ln
\frac{n}{p_s} - \delta_{n}} = \frac{p_s}{n} e^{- \delta_{n}} \ln
\frac{n}{p_s}, \la{mxrn2}
\end{align}
and with $\Delta$ denoting $\pi {r_n}^2 p_s n$ (note that $\Delta=
\Theta(\ln n)$ from Lemma \ref{lems}),
\begin{align}
r_n & = \pi^{-\frac{1}{2}}{p_s}^{-\frac{1}{2}}n^{-\frac{1}{2}}
\Delta^{\frac{1}{2}}.  \la{mxrn}
\end{align}
Then using (\ref{mxrn2}) and (\ref{mxrn}) in (\ref{pixlo}), \f
\begin{align}
  &  n\mathbb{P}[I_x] \nonumber \\
   & \hspace{-2pt} \geq
n \hspace{-1pt} \cdot \hspace{-1pt} 2
\pi^{\frac{1}{2}}{p_s}^{-\frac{1}{2}}n^{-\frac{1}{2}}\Delta^{
-\frac{1}{2}} \hspace{-2pt}\cdot\hspace{-2pt}
p_s^{\frac{1}{2}}{n^{-\frac{1}{2}}} e^{-\frac{\delta_n}{2}}
\bigg(\ln \frac{n}{p_s}\bigg)^{\frac{1}{2}}
\hspace{-2pt}\cdot\hspace{-2pt}
 [1\hspace{-1pt}-\hspace{-1pt}o(1)]  \nonumber \\
& \hspace{-2pt} \geq 2 \pi^{\frac{1}{2}}
     \bigg(\Delta^{-1}\ln \frac{n}{p_s}\bigg)^{\frac{1}{2}} e^{-\frac{\delta_n}{2}}
   \cdot [1\hspace{-1pt}-\hspace{-1pt}o(1)]. \nonumber
 \end{align}
From $\Delta = \Theta(\ln n)$ and $ \ln \frac{n}{p_s} = \Theta(\ln
n)$ in Lemma \ref{lems}, with $\lim\limits_{n\to \infty}\alpha_n= -
  \infty$ producing $\lim\limits_{n\to \infty}\delta_n= -
  \infty$, we have
\begin{align}
  & \lim_{n\to \infty} \big\{ n\mathbb{P}[I_x] \big\} = \infty.\nonumber
 \end{align}
%
%
%

With $p_s =  O \left( \frac{1}{n^{1/3}\ln n} \right)$, by
\cite[Equation (8.21)]{Penrose2013}, \f
\begin{align}
  & \lim_{n\to \infty} \big\{ n\mathbb{P}[I_x] \big\} = \infty\textrm{ if }\lim_{n\to \infty}\delta_n= -
  \infty.\nonumber
 \end{align}

\subsubsection{Establishing (\ref{PIxIy}) on the unit square $\mathcal {S}$}

Clearly, (\ref{proIxIy}) and (\ref{leq_IxIy}) still hold. Here we
consider the network area $\mathcal {A}$ as the unit square
$\mathcal {S}$. For any $\hat{v_x} \in \mathcal {S}$ and any
$\hat{v_y} \in \mathcal {S}$, we have $|D_{r_n}(\hat{v_x}) \bcap
D_{r_n}(\hat{v_y})| \leq \pi {r_n}^{2}$,
which is applied to (\ref{leq_IxIy}) so that%
%
\begin{align}
  & \mathbb{P}\big[\hspace{1pt}I_x
\bcap I_y \boldsymbol{\mid} (|S_{xy}| = u)\hspace{1pt}\big]
\nonumber \\   &
  \leq
e^{ \pi {r_n}^{2} \phi_u n}  \int_{\mathcal {S}} e^{-n p_s
|D_{{r_n}}(\hat{v_x})|}
 \, \textrm{d}\hat{v_x}  \int_{\mathcal {S}} e^{-n p_s
|D_{{r_n}}(\hat{v_y})|} \, \textrm{d}\hat{v_y} \nonumber \\
& = e^{ \pi {r_n}^{2} \phi_u n}  \cdot
\big\{\mathbb{P}[I_x]\big\}^2, \label{ev_pxysqr}
\end{align}
where we use the result that $\int_{\mathcal {S}} e^{-n p_s
|D_{{r_n}}(\hat{v_x})|}
 \, \textrm{d}\hat{v_x} $ and $ \int_{\mathcal {S}} e^{-n p_s
|D_{{r_n}}(\hat{v_y})|} \, \textrm{d}\hat{v_y} $ both equal $
\mathbb{P}[I_x]$ in the last step of (\ref{ev_pxysqr}). Then using
(\ref{ev_pxysqr}) in (\ref{proIxIy}), we obtain
\begin{align}
\hspace{-3pt} \mathbb{P}[I_x \bcap I_y] \hspace{-1pt} \leq
\hspace{-1pt} \big\{\mathbb{P}[I_x]\big\}^2 \hspace{-1pt} \cdot
\hspace{-1pt} \sum_{u=0}^{K_n}\Big\{ \mathbb{P}[
|S_{xy}|\hspace{-1pt} = \hspace{-1pt}u ] e^{ \pi {r_n}^{2} \phi_u n}
\Big\}. \label{mathbbIxIysqr}
\end{align}
Since (\ref{e2pir}) also holds here, we know from (\ref{e2pir}) and
(\ref{mathbbIxIysqr}) that the proof of (\ref{PIxIy}) on $\mathcal
{S}$ is completed once we prove
\begin{align}
2\pi {r_n}^{2}n \cdot \frac{{K_n}^4}{{P_n}^2} +
\frac{{K_n}^2}{P_n-K_n} \cdot e^{ \pi {r_n}^{2}n\frac{K_n}{P_n}} & =
o(1). \label{pdf}
\end{align}
Under (\ref{thm:s_Kn}) with $\nu_n = o(1)$, we have
$\frac{{K_n}^2}{P_n} \leq \frac{{\nu_n}^2}{\ln n} =
o\big(\frac{1}{\ln n}\big)$, so (\ref{pssim}) holds. From Lemma
\ref{lems}, it is always true that $\pi
{r_n}^{2}n\cdot\frac{{K_n}^2}{P_n} = \Theta(\ln n)$, which along
with $\frac{{K_n}^2}{P_n} = o\big(\frac{1}{\ln n}\big)$ and
condition $ K_n \geq c_2 \sqrt{\frac{P_n \ln n}{n^{c_3}}}$ in
(\ref{thm:s_Kn}) with $c_3>0$ leads to $\pi {r_n}^{2}n \cdot
\frac{{K_n}^4}{{P_n}^2} = o(1)$ and
\begin{align}
 e^{\pi {r_n}^{2}n\frac{K_n}{P_n}} =  e^{\pi
{r_n}^{2}n\frac{{K_n}^2}{P_n}\cdot {K_n}^{-1}} \to 1,\textrm{ as }n
\to \infty.
\end{align}
Then (\ref{pdf}) is proved, completing the proof of (\ref{PIxIy}) on
$\mathcal {S}$.

\section{ESTABLISHING THE ONE--LAWS}

%

From (\ref{thm:s_Kn}), we have $K_n = \Omega\left((\ln n)^3\right)$,
$\frac{K_n}{P_n} = O\left(\frac{1}{n \ln n}\right) $ and
$\frac{{K_n}^2}{P_n} = O\left(\frac{1}{ \ln n}\right)$.
 Therefore, in view that all
conditions of Lemma \ref{cp_rig_er} are satisfied, and considering
that connectivity is a monotone increasing graph property, we apply
Lemma \ref{cp_rig_er} to obtain for some $p_n$ with (\ref{eq_pn}),
\begin{align}
& \hspace{-1pt} \mathbb{P}[ G_{RKG}(n,
\hspace{-0.2pt}K_n,\hspace{-0.2pt} P_n) \hspace{-1.7pt} \bcap
\hspace{-1pt} G_{RGG}(n, \hspace{-0.2pt}r_n,
\hspace{-0.2pt}\mathcal{S}) \textrm{ is connected.} ] \nonumber
\\ &  \hspace{-2.9pt} \geq \hspace{-1pt}
 \mathbb{P}[ G_{ER}(n,\hspace{-0.2pt} p_n\hspace{-0.1pt}) \hspace{-1.7pt}
  \bcap \hspace{-1pt} G_{RGG}(n, \hspace{-0.2pt}r_n,\hspace{-0.2pt}
\mathcal{S}) \textrm{ is connected.} ] \hspace{-1pt}- \hspace{-1pt}
o(1).
 \end{align}

 Here we also define $\nu_n$ through (\ref{lem:erg_rgg_rnpn}). We will show that under (\ref{thm:s_Kn}) and (\ref{thm:s:rnKnPn}),
$\nu_n$ specified in (\ref{lem:erg_rgg_rnpn}) of Lemma
\ref{erg_rgg_conn} equals $\alpha_n \pm O(1)$, where $\alpha_n$ is
set in (\ref{thm:s:rnKnPn}).

In order to assess $\nu_n$, we see from (\ref{lem:erg_rgg_rnpn})
that it is useful to evaluate $\ln \frac{1}{p_n}$ and $ \ln \ln
\frac{1}{p_n}$. Given (\ref{eq_pn}), we obtain
\begin{align}
 \ln \frac{1}{p_n} & =\ln \frac{P_n}{{K_n}^2} - \ln \left[1 -
 O \left(\frac{1}{ \ln n}\right)\right] \nonumber \\ &
  =\ln \frac{P_n}{{K_n}^2} -
 O \left(\frac{1}{ \ln n}\right), \label{ln1pn}
\end{align}
and with $\frac{{K_n}^2}{P_n} = O\left(\frac{1}{ \ln n}\right)$,
\begin{align}
 \ln \ln \frac{1}{p_n} & = \ln \left[\ln \frac{P_n}{{K_n}^2} -
 O \left(\frac{1}{ \ln n}\right)\right] \nonumber \\ &
 = \ln \left\{\ln \frac{P_n}{{K_n}^2} \left[ 1 -
 O \left(\frac{1}{ \ln n\ln \ln n}\right)\right] \right\} \nonumber \\ &
 = \ln \ln \frac{P_n}{{K_n}^2} -
 O \left(\frac{1}{ \ln n \ln \ln n}\right).  \label{lnln1pn}
\end{align}

By (\ref{eq_pn}), it holds that
\begin{align}
 p_n &  \sim \frac{{K_n}^2}{P_n},\nonumber
\end{align}
so we have
\begin{align}
 p_n & \hspace{-.6pt}=\hspace{-.6pt} \omega\hspace{-1pt}
 \left(\hspace{-1pt} \frac{1}{n^{1/3}\ln n} \hspace{-1pt}\right)
 \textrm{\hspace{-2.7pt} if and only if \hspace{-1pt}}
 \frac{{K_n}^2}{P_n} \hspace{-.6pt}=\hspace{-.6pt} \omega\hspace{-1pt}
  \left(\hspace{-1pt} \frac{1}{n^{1/3}\ln n} \hspace{-1pt}\right)\hspace{-1pt},
  \label{pnpn1}
\end{align}
and
\begin{align}
 p_n & \hspace{-.6pt}=\hspace{-.6pt} O\hspace{-1pt}
 \left(\hspace{-1pt} \frac{1}{n^{1/3}\ln n} \hspace{-1pt}\right)
 \textrm{\hspace{-2.7pt} if and only if \hspace{-1pt}}
 \frac{{K_n}^2}{P_n} \hspace{-.6pt}=\hspace{-.6pt} O\hspace{-1pt}
  \left(\hspace{-1pt} \frac{1}{n^{1/3}\ln n} \hspace{-1pt}\right)\hspace{-1pt}.
  \label{pnpn2}
\end{align}

Now it is ready to compute $\nu_n$ according to
(\ref{lem:erg_rgg_rnpn}). On the one hand, for $\frac{{K_n}^2}{P_n}
= \omega \left( \frac{1}{n^{1/3}\ln n} \right)$ which is equivalent
to $p_n = \omega \left( \frac{1}{n^{1/3}\ln n} \right)$ in view of
(\ref{pnpn1}), we apply (\ref{lem:erg_rgg_rnpn}) (\ref{ln1pn}) and
(\ref{lnln1pn}) to derive
\begin{align}
 \nu_n & = \pi {r_n}^2 p_n \cdot n - \bigg(\ln \frac{n}{p_n} - \ln \ln
\frac{n}{p_n}\bigg)  \nonumber \\ & = \pi {r_n}^2
\cdot\frac{{K_n}^2}{P_n} \cdot \left[1 -
 O \left(\frac{1}{ \ln n}\right)\right]  \cdot n \nonumber \\ &
 \quad - \bigg[
\ln \frac{n P_n}{{K_n}^2}  -
 O \left(\frac{1}{ \ln n}\right) \bigg] \nonumber \\ &
 \quad + \ln \ln \frac{nP_n}{{K_n}^2} -
 O \left(\frac{1}{ \ln n \ln \ln n}\right)
\end{align}
With $\pi {r_n}^2 \cdot\frac{{K_n}^2}{P_n} = \Theta\big(\frac{\ln
n}{n}\big)$,
\begin{align}
 \nu_n & =
\pi {r_n}^2 \hspace{-1pt}\cdot\hspace{-1pt}\frac{{K_n}^2}{P_n}
\hspace{-1pt}\cdot\hspace{-1pt} n \hspace{-.5pt}-\hspace{-.5pt} \ln
\frac{n P_n}{{K_n}^2} \hspace{-.5pt}+\hspace{-.5pt}
\ln \ln \frac{nP_n}{{K_n}^2} \nonumber \\
& \quad - \hspace{-.5pt}\Theta(\ln n)
\hspace{-1pt}\cdot\hspace{-1pt} O \left(\hspace{-1pt}\frac{1}{ \ln
n}\hspace{-1pt}\right) \hspace{-.5pt}+\hspace{-.5pt} O
\left(\hspace{-1pt}\frac{1}{ \ln n}\hspace{-1pt}\right)
\hspace{-.5pt}-\hspace{-.5pt}
 O \left(\hspace{-1pt}\frac{1}{ \ln n \ln \ln n}\hspace{-1pt}\right)
 \nonumber \\ &
  = \alpha_n \hspace{-.5pt}\pm\hspace{-.5pt} O(1) .\label{beta1}
  \end{align}

%

On the other hand, for $\frac{{K_n}^2}{P_n} = O \left(
\frac{1}{n^{1/3}\ln n} \right)$ that is equivalent to $p_n = O
\left( \frac{1}{n^{1/3}\ln n} \right)$ in view of (\ref{pnpn2}), we
use (\ref{lem:erg_rgg_rnpn}) (\ref{ln1pn}) and (\ref{lnln1pn}) to
obtain
\begin{align}
 \nu_n & = \pi {r_n}^2 p_n \cdot n - \bigg( 4\ln \frac{1}{p_n} - 4\ln \ln
\frac{1}{p_n} \bigg)  \nonumber \\ & = \pi {r_n}^2
\cdot\frac{{K_n}^2}{P_n} \cdot \left[1 -
 O \left(\frac{1}{ \ln n}\right)\right]  \cdot n \nonumber \\ &
 \quad - 4\bigg[\ln \frac{P_n}{{K_n}^2} -
 O \left(\frac{1}{ \ln n}\right) \bigg] \nonumber \\ &
 \quad + 4\bigg[\ln \ln \frac{P_n}{{K_n}^2} -
 O \left(\frac{1}{ \ln n \ln \ln n}\right) \bigg]
\end{align}
Then
\begin{align}
 \nu_n & =
\pi {r_n}^2 \hspace{-1pt}\cdot\hspace{-1pt}\frac{{K_n}^2}{P_n}
\hspace{-1pt}\cdot\hspace{-1pt} n \hspace{-.5pt}-\hspace{-.5pt} 4\ln
\frac{ P_n}{{K_n}^2} \hspace{-.5pt}+\hspace{-.5pt}
4\ln \ln \frac{ P_n}{{K_n}^2} \nonumber \\
& \quad - \hspace{-.5pt}\Theta(\ln n)
\hspace{-1pt}\cdot\hspace{-1pt} O \left(\hspace{-1pt}\frac{1}{ \ln
n}\hspace{-1pt}\right) \hspace{-.5pt}+\hspace{-.5pt} O
\left(\hspace{-1pt}\frac{1}{ \ln n}\hspace{-1pt}\right)
\hspace{-.5pt}-\hspace{-.5pt}
 O \left(\hspace{-1pt}\frac{1}{ \ln n \ln \ln n}\hspace{-1pt}\right)
 \nonumber \\ &
  = \alpha_n \hspace{-.5pt}\pm\hspace{-.5pt} O(1). \label{beta2}
  \end{align}

Summarizing (\ref{beta1}) and (\ref{beta2}), with (\ref{thm:s_Kn})
and (\ref{thm:s:rnKnPn}), $\nu_n$ defined in
(\ref{lem:erg_rgg_rnpn}) equals $\alpha_n \pm O(1)$ specified in
(\ref{thm:s:rnKnPn}). Then by Lemmas \ref{erg_rgg_conn} and
\ref{cp_rig_er}, the result follows.

\section{ESTABLISHING THE LEMMAS} \label{sec:prf:lemmas}


%
%
%

\subsection{The Proof of Lemma \ref{lem:IxIxIy}} \label{sec:prf:lem:IxIxIy}

When node $v_x$ is at position $\hat{v_x}$, the number of nodes
within area $D_{r_n}(\hat{v_x})$ follows a Poisson distribution with
mean $n D_{r_n}(\hat{v_x})$; and to have an edge with $v_x$ in graph
$G_{\textrm{Poisson}}(n, \theta_n, \mathcal{A})$, a node not only
has to be within a $D_{r_n}(\hat{v_x})$ but also has to share at
least a key with node $v_x$. Then by Lemma \ref{lem:palm}, the
number of nodes neighboring to $v_x$ at $\hat{v_x}$ follows a
Poisson distribution with mean $n p_s |D_{r_n}(\hat{v_x})|$; and the
probability that such number is $0$ equals $e^{-n p_s
|D_{r_n}(\hat{v_x})|}$. Integrating $\hat{v_x}$ over $\mathcal{A}$,
we derive the probability that
  node $v_x$ is isolated (i.e., $\mathbb{P}[I_x]$) via
\begin{align}
 \mathbb{P}[I_x]  &= n \int_{ \mathcal{A} }
 e^{-n p_s |D_{r_n}(\hat{v_x})|} \, \textrm{d}\hat{v_x}; \nonumber
\end{align}
namely, (\ref{eq_Ix}) follows.

Below we demonstrate (\ref{eq_IxIy_Sxyu}) and (\ref{eq_IxIy_Sxy0}).
For the ease of explanation, we define $$\mathcal
{E}\big[(v_x\textrm{ at }\hat{v_x}) \bcap (v_y\textrm{ at
}\hat{v_y}) \bcap (|S_{xy}| = u)\big]$$ as the event that
\begin{itemize}
  \item nodes $v_x$ and $v_y$ are at positions
$\hat{v_x}$ and $\hat{v_y}$, respectively;
  \item and $v_x$ and $v_y$ share a certain
number $u$ of keys, where $u = 0, 1, \ldots, K_n$.
\end{itemize}
Conditioning on $\mathcal {E}\big[(v_x\textrm{ at }\hat{v_x}) \bcap
(v_y\textrm{ at }\hat{v_y}) \bcap (|S_{xy}| = u)\big]$, we further
define $N(\mathcal {E})$ as the number of nodes different from $v_x$
and $v_y$, and neighboring to at least one of $v_x$ and $v_y$. By
Lemma \ref{lem:palm}, $N(\mathcal {E})$ follows a Poisson
distribution with mean
\begin{align}
n \cdot {\mathbb{P}}\bigg[\hspace{1pt}\textrm{E}_{{x} j} \bcup
\textrm{E}_{{y}
    j}  \hspace{2pt}\boldsymbol{\bigg|}\hspace{2pt} { (|S_{xy}| = u)} \begin{array}{l} {\small \selectfont \bcap (v_x\textrm{ is at }\hat{v_x}),}
 \\ {\small \selectfont \bcap (v_y\textrm{ is at }\hat{v_y}).} \end{array}
  \bigg] ,\label{poxory}
\end{align}
which we denote by $\lambda_{\hat{v_x}, \hat{v_y}, u}$ below.

Conditioning on $\mathcal {E}\big[(v_x\textrm{ at }\hat{v_x})
\hspace{-1pt} \bcap \hspace{-1pt} (v_y\textrm{ at }\hat{v_y})
\hspace{-1pt} \bcap \hspace{-1pt} (|S_{xy}| \hspace{-1pt} =
\hspace{-1pt} u)\big]$, event $I_x \bcap I_y$ (i.e., the event that
nodes $v_x$ and $v_y$ are both isolated) is equivalent to
$N(\mathcal {E}) \bcap \overline{E_{xy}}$. Conditioning on event
$(|S_{xy}| = u)$, for event $\overline{E_{xy}}$ to occur, the
distance between $v_x$ at $\hat{v_x}$ and $v_y$ at $\hat{v_y}$ has
to be greater than distance $r_n$ for $u=1,2,\ldots,K_n$; and there
is no such requirement for $u=0$ as $(|S_{xy}| = 0)$ already implies
$\overline{E_{xy}}$. Therefore, we obtain
\begin{align}
& \mathbb{P}\Big[I_x \bcap I_y
\hspace{2pt}\boldsymbol{\Big|}\hspace{2pt} \mathcal
{E}\big[(v_x\textrm{ at }\hat{v_x}) \bcap (v_y\textrm{ at
}\hat{v_y}) \bcap (|S_{xy}| = 0)\big]\Big]
\nonumber \\
& \quad  = \mathbb{P}\big[N(\mathcal {E}) = 0 \big]  \nonumber \\
& \quad  =
  e^{-n\lambda_{\hat{v_x}, \hat{v_y}, 0}}, \label{xy1}
\end{align}
where $\lambda_{\hat{v_x}, \hat{v_y}, 0}$ stands for
$\lambda_{\hat{v_x}, \hat{v_y}, u}$ when $u=0$; and for
$u=1,2,\ldots,K_n$,
\begin{itemize}
  \item if $\hat{v_x}$ and $\hat{v_y}$ has a distance greater than
$r_n$, then
\begin{align}
& \mathbb{P}\Big[I_x \bcap I_y
\hspace{2pt}\boldsymbol{\Big|}\hspace{2pt} \mathcal
{E}\big[(v_x\textrm{ at }\hat{v_x}) \bcap (v_y\textrm{ at
}\hat{v_y}) \bcap (|S_{xy}| = u)\big]\Big]
\nonumber \\
& \quad  = \mathbb{P}\big[N(\mathcal {E}) = 0 \big] \nonumber \\
& \quad  =
  e^{-n\lambda_{\hat{v_x}, \hat{v_y}, u}}; \label{xy2}
\end{align}
  \item and if $\hat{v_x}$ and $\hat{v_y}$ has a distance no greater than
distance $r_n$, then
\begin{align}
& \mathbb{P}\Big[I_x \bcap I_y
\hspace{2pt}\boldsymbol{\Big|}\hspace{2pt} \mathcal
{E}\big[(v_x\textrm{ at }\hat{v_x}) \bcap (v_y\textrm{ at
}\hat{v_y}) \bcap (|S_{xy}| = u)\big]\Big]\nonumber \\
& \quad   =0. \label{xy3}
\end{align}
\end{itemize}
For $u=0,1,\ldots,K_n$, integrating $$\mathbb{P}\Big[I_x \bcap I_y
\hspace{2pt}\boldsymbol{\Big|}\hspace{2pt} \mathcal
{E}\big[(v_x\textrm{ at }\hat{v_x}) \bcap (v_y\textrm{ at
}\hat{v_y}) \bcap (|S_{xy}| = u)\big]\Big]$$ with $\hat{v_x}$ over
$\mathcal {A}$ and $\hat{v_y}$ also over $\mathcal {A}$, we then
obtain $\mathbb{P}\big[I_x \bcap I_y \boldsymbol{\mid} (|S_{xy}| =
u)\big]$. Hence, in view of (\ref{xy1}--\ref{xy3}), it is easy to
establish
\begin{align}
& \mathbb{P}\big[I_x \bcap I_y \boldsymbol{\mid} (|S_{xy}| = u)\big]
\nonumber \\
& ~  = \int_{\mathcal {A}} \int_{\mathcal {A}\setminus
D_r(\hat{v_x})}
  e^{-n\lambda_{\hat{v_x}, \hat{v_y}, u}}
  \, \textrm{d}\hat{v_x} \, \textrm{d}\hat{v_y}; \label{prIxIyu}
\end{align}
and
\begin{align}
& \mathbb{P}\big[I_x \bcap I_y \boldsymbol{\mid} (|S_{xy}| = 0)\big]
\nonumber \\
& ~  = \int_{\mathcal {A}} \int_{\mathcal {A}}
  e^{-n\lambda_{\hat{v_x}, \hat{v_y}, 0}}
  \, \textrm{d}\hat{v_x} \, \textrm{d}\hat{v_y}. \label{prIxIy0}
\end{align}
 To evaluate (\ref{prIxIyu}) and (\ref{prIxIy0}), we calculate
$\lambda_{\hat{v_x}, \hat{v_y}, u}$ below based on its expression in
(\ref{poxory}). By (\ref{poxory}), it is clear that
\begin{align}
& \lambda_{\hat{v_x}, \hat{v_y}, u} \nonumber \\  & = n \cdot
{\mathbb{P}}\bigg[\hspace{1pt}\textrm{E}_{{x} j}
\hspace{2pt}\boldsymbol{\bigg|}\hspace{2pt} { (|S_{xy}| = u)}
\begin{array}{l} {\small \selectfont \bcap (v_x\textrm{ is at
}\hat{v_x}),}
 \\ {\small \selectfont \bcap (v_y\textrm{ is at }\hat{v_y}).} \end{array}
  \bigg]  \nonumber \\ & \quad +
 n \cdot {\mathbb{P}}\bigg[\hspace{1pt}\textrm{E}_{{x} j} \hspace{2pt}\boldsymbol{\bigg|}\hspace{2pt} { (|S_{xy}| = u)} \begin{array}{l} {\small \selectfont \bcap (v_x\textrm{ is at }\hat{v_x}),}
 \\ {\small \selectfont \bcap (v_y\textrm{ is at }\hat{v_y}).} \end{array}
  \bigg] \nonumber \\ &  \quad
- n \cdot {\mathbb{P}}\bigg[\hspace{1pt}\textrm{E}_{{x} j} \bcap
\textrm{E}_{{y}
    j}  \hspace{2pt}\boldsymbol{\bigg|}\hspace{2pt} { (|S_{xy}| = u)} \begin{array}{l} {\small \selectfont \bcap (v_x\textrm{ is at }\hat{v_x}),}
 \\ {\small \selectfont \bcap (v_y\textrm{ is at }\hat{v_y}).} \end{array}
  \bigg]. \label{lxyu}
\end{align}
To further assess (\ref{lxyu}), we have the following observations.
To begin with,
\begin{align}
\textrm{E}_{{x} j} & = \textrm{K}_{{x} j} \cap
\big[\hspace{1pt}v_j\textrm{ is within }
  D_{r}(\hat{v_x}) \hspace{1pt}\big]. \label{eij_def}
\end{align}
Each of $K_{{x} j}$ and $K_{{y} j}$ is independent of $(|S_{xy}| =
u)$, but $K_{{x} j} \bcap K_{{y} j}$ is not independent of
$(|S_{xy}| = u)$. Clearly,
\begin{align}
 {\mathbb{P}}\big[ K_{{x} j}
 \boldsymbol{\mid}  (|S_{xy}| = u) \big]  &  =
 {\mathbb{P}}\big[ K_{{x} j} \big] = p_s, \label{e1}
 \end{align}
 and
\begin{align}
  {\mathbb{P}}\big[ K_{{y} j}
 \boldsymbol{\mid}  (|S_{xy}| = u) \big]  &  =
 {\mathbb{P}}\big[ K_{{y} j} \big] = p_s.\label{e2}
\end{align}
We also recall
\begin{align}
 \phi_u = {\mathbb{P}}\big[\hspace{1pt} K_{{x} j} \bcap K_{{y} j}
\boldsymbol{\mid} (|S_{xy}| = u) \hspace{1pt}\big].
\label{phi_u_rec}
\end{align}
Then we use (\ref{eij_def}--\ref{phi_u_rec}) in (\ref{lxyu})
 to derive
\begin{align}
\lambda_{\hat{v_x}, \hat{v_y}, u}
    & = n p_s \cdot \mathlarger{\big[} \big|D_{r_n}(\hat{v_x}) \big|
 + \big|D_{r_n}(\hat{v_y}) \big| \mathlarger{\big]}  \nonumber \\
 & \quad  - n \phi_u \cdot
\big|D_{r_n}(\hat{v_x}) \bcap
 D_{r_n}(\hat{v_y})\big|. \label{lambdasum}
\end{align}
Substituting (\ref{lambdasum}) into (\ref{prIxIyu}) and
(\ref{prIxIy0}), we establish (\ref{eq_IxIy_Sxyu}) and
(\ref{eq_IxIy_Sxy0}), respectively. \qedb

\subsection{The Proof of Lemma \ref{lemt}}

Given (\ref{thm:t:rnKnPn}) and $\alpha_{n} < 0$, we obtain ${
\pi{r_n}^2 \cdot \frac{{K_n}^2}{P_n}} \cdot n \leq \ln n$, which
together with $\frac{{K_n}^2}{P_n} \geq  \frac{{\mu_n}^2\ln n}{n}
=\omega\big(\frac{\ln n}{n}\big) $ resulted from condition
(\ref{thm:t_Kn}) brings about $r_n = o(1)$ because of $\pi r_{n}^2 =
O\big(\frac{\ln n}{n}\big) \cdot
 \big[\omega\big(\frac{\ln n}{n}\big)\big]^{-1}  =o(1)$.
 With
condition (\ref{thm:t_Kn}), \f $\frac{{K_n}^2}{P_n} \leq
\frac{{c_1}^2}{\ln n} = O(\frac{1}{\ln n})$.  Then from
(\ref{pssim}) (\ref{thm:t:rnKnPn}) and ${ \pi{r_n}^2 \cdot
\frac{{K_n}^2}{P_n}} \cdot n \leq \ln n$, we obtain
\begin{align}
\pi {r_n}^2 {p_s} n 
 & =  \pi {r_n}^2 \cdot \frac{{K_n}^2}{P_n} \cdot n \cdot
\bigg[1-O\bigg(\frac{{K_n}^2}{P_n}\bigg)\bigg] \nonumber \\
 & = \ln n +
 \alpha_{n} - O(\ln n) \cdot O\bigg(\frac{1}{\ln n}\bigg) \nonumber \\
 & =\ln n + \alpha_{n} - O(1). \nonumber
 \end{align}

\subsection{The Proof of Lemma \ref{lems}} \vspace{-2pt}

From $\frac{{c_2}^2\ln n}{n^{c_3}}  \leq \frac{{K_n}^2}{P_n}
 \leq \frac{{\nu_n}^2}{\ln n} $, (\ref{psleqKnPn}) and (\ref{pssim}), there exists
$\varphi_n$ with $\lim\limits_{n \to \infty}\varphi_n = 1$ such that 
 $\frac{{c_2}^2\ln n}{n^{c_3}} \cdot \varphi_n  \leq p_s
 \leq \frac{{\nu_n}^2}{\ln n} $. Then $\ln (n / p_s) \leq (1+c_3) \ln n - \ln \ln n -\ln \varphi_n  - 2\ln c_2$
and $\ln (n / p_s)  \geq   \ln n + \ln \ln n -2 \ln \nu_n$ hold,
leading to $\ln (n / p_s)  = \Theta(\ln n)$. Similarly, $\ln
\frac{nP_n}{{K_n}^2}  = \Theta(\ln n)$. For $\frac{{K_n}^2}{P_n} = O
\left( \frac{1}{n^{1/3}\ln n} \right)$, then in view of $\ln
\frac{nP_n}{{K_n}^2}  = \Theta(\ln n)$, $\pi {r_n}^2 \cdot
\frac{{K_n}^2}{P_n} \cdot n = \ln \frac{n P_n}{{K_n}^2} - \ln \ln
\frac{n P_n}{{K_n}^2} + \alpha_{n}$ resulted from
(\ref{thm:s:rnKnPn}), and $ |\alpha_n| = O(\ln \ln n)$, we obtain
$\pi {r_n}^2 \cdot\frac{{K_n}^2}{P_n} = \Theta\big(\frac{\ln
n}{n}\big)$. On the other hand, for $\frac{{K_n}^2}{P_n} = \omega
\left( \frac{1}{n^{1/3}\ln n} \right)$, then we apply $\ln
\frac{P_n}{{K_n}^2} = \ln\left[\omega \left(n^{1/3}\ln
n\right)\right] = \Theta(\ln n)$ and condition $\pi {r_n}^2 \cdot
\frac{{K_n}^2}{P_n} \cdot n 4\ln \frac{P_n}{{K_n}^2} - 4\ln \ln
 \frac{P_n}{{K_n}^2} + \alpha_{n}$ from
(\ref{thm:s:rnKnPn}) and $ |\alpha_n| = O(\ln \ln n)$ to derive $\pi
{r_n}^2 \cdot\frac{{K_n}^2}{P_n} = \Theta\big(\frac{\ln n}{n}\big)$.
Hence, under either $\frac{{K_n}^2}{P_n} = O \left(
\frac{1}{n^{1/3}\ln n} \right)$ or $\frac{{K_n}^2}{P_n} = \omega
\left( \frac{1}{n^{1/3}\ln n} \right)$, it holds that $\pi {r_n}^2
\cdot\frac{{K_n}^2}{P_n} = \Theta\big(\frac{\ln n}{n}\big)$, which
along with (\ref{pssim}) and $\frac{{K_n}^2}{P_n} \geq
\frac{{c_2}^2\ln n}{n^{c_3}} $ resulting in $\pi {r_n}^2 \cdot p_s =
\Theta\big(\frac{\ln n}{n}\big)$ and $r_n  = o(1)$, respectively.

In order to assess $\delta_n$, we see from (\ref{deltan}) that it is
useful to evaluate $\ln \frac{1}{p_s}$ and $ \ln \ln \frac{1}{p_s}$.
From $\frac{{K_n}^2}{P_n}
 \leq \frac{{\nu_n}^2}{\ln n} $ and (\ref{pssim}), then we obtain $\ln \frac{1}{p_s}
 = \ln \frac{P_n}{{K_n}^2} -
 O \left(\frac{1}{ \ln n}\right)$, which leads to\\
  $\ln \ln \frac{1}{p_s} \hspace{-2pt}= \hspace{-2pt}\ln\big[ \ln \frac{P_n}{{K_n}^2} -
 O \left(\frac{1}{ \ln n}\right)\big]\hspace{-2pt}= \hspace{-2pt}\ln \ln \frac{P_n}{{K_n}^2} -
 O \left(\frac{1}{ \ln n \ln \ln n}\right)$.\vspace{2pt}


Now it is ready to compute $\delta_n$. In view of (\ref{pssim}),
clearly $p_s = O \left( \frac{1}{n^{1/3}\ln n} \right)$ (resp., $p_s
= \omega \left( \frac{1}{n^{1/3}\ln n} \right)$) is equivalent to
$p_s = \omega \left( \frac{1}{n^{1/3}\ln n} \right)$ (resp.,
$\frac{{K_n}^2}{P_n} = \omega \left( \frac{1}{n^{1/3}\ln n}
\right)$). We apply the expressions of $\ln \frac{1}{p_s}$ and $\ln
\ln \frac{1}{p_s}$ above, $\pi {r_n}^2 \cdot\frac{{K_n}^2}{P_n}
\cdot n = \Theta(\ln n)$, (\ref{thm:s:rnKnPn}) and (\ref{deltan}) to
show that 
 under either $p_s = O \left( \frac{1}{n^{1/3}\ln n} \right)$ or
 $p_s = \omega \left( \frac{1}{n^{1/3}\ln n} \right)$, we always have\vspace{-5pt}
\begin{align}
 \delta_n & \hspace{-2pt}= \hspace{-2pt}
\alpha_n \hspace{-2pt} - \hspace{-2pt}\Theta(\ln n)
\hspace{-1pt}\cdot\hspace{-1pt} O \left(\hspace{-1pt}\frac{1}{ \ln
n}\hspace{-1pt}\right) \hspace{-2pt}+\hspace{-2pt} O
\left(\hspace{-1pt}\frac{1}{ \ln n}\hspace{-1pt}\right)
\hspace{-2pt}-\hspace{-2pt}
 O \left(\hspace{-1pt}\frac{1}{ \ln n \ln \ln n}\hspace{-1pt}\right)
 \nonumber \vspace{-5pt} \\ &
  \hspace{-2pt}=\hspace{-2pt} \alpha_n \hspace{-1.5pt}\pm\hspace{-1.5pt} O(1)
  . \nonumber
  \end{align}
~\vspace{-17pt}
%

%
%

\subsection{The Proof of Lemma \ref{lemu}} \vspace{-2pt}

To begin with, it holds that \vspace{-3pt}
\begin{align}
& \mathbb{P}[K_{{ xz}}\cap K_{ yz} \boldsymbol{\mid} (|S_{xy}| = u)]
\nonumber \\   & ~= \mathbb{P}[K_{{xz}} \boldsymbol{\mid} (|S_{xy}|
= u)] + \mathbb{P}[
K_{yz} \boldsymbol{\mid} (|S_{xy}| = u)] \nonumber \\
& ~\quad- ( 1- \mathbb{P}[\overline{K_{{ xz}}}\cap \overline{K_{
yz}}\boldsymbol{\mid} (|S_{xy}| = u)])
 \nonumber \\  &~ = 2 p_s -   1 + \binom{P_n -
(2K_n -u)}{K_n}\bigg/\binom{P_n}{K_n} . \vspace{-5pt} \label{m1}
\end{align}
By {\cite[Lemma 7.1]{yagan}} and {\cite[Fact 2]{ZhaoYaganGligor}},
\f
\begin{align}
&\binom{P_n - (2K_n -u)}{K_n}\bigg/\binom{P_n}{K_n} \leq
\bigg(1-\frac{2K_n -u}{P_n}\bigg)^{K_n} \nonumber \\  &
   \leq
 1-\frac{(2K_n -u)K_n}{P_n}\cdot{K_n} + \frac{1}{2} \bigg[
 \frac{(2K_n -u)K_n}{P_n}\bigg]^2. \vspace{-5pt} \label{m2}
\end{align}
Applying (\ref{psleqKnPn}) (i.e., $p_{s}  \leq \frac{{K_n}^2}{P_n}$)
and (\ref{m2}) to (\ref{m1}), we have \vspace{-3pt}
\begin{align}
\mathbb{P}[({K}_{xz} \cap {K}_{yz} \boldsymbol{\mid} (|S_{xy}| = u)]
& \leq
 \frac{ u{K_n}}{P_n} + \frac{2{K_n}^4}{{P_n}^2}.
\nonumber
\end{align}


%
%
%

%

\subsection{The Proof of Lemma \ref{Poissonization1}}

From (\ref{prIxe}), \f
\begin{align}
n \mathbb{P}[I_x] & = n e^{-\pi {r_n}^2
 p_s n} = n e^{-\ln n - \alpha_n \pm O(1)}. \nonumber
\end{align}
With $|\alpha_n| = O(\ln \ln n)$, we have for any constant
$\epsilon>0$,
\begin{align}
  & n\mathbb{P}[I_x] = o(n^{\epsilon}) .  \nonumber
 \end{align}

\subsection{The Proof of Lemma \ref{Poissonization2}}

As in Section \ref{sec:sq}, we partition $\mathcal{S}$ according to
Figure
 \ref{partition} and define $T_i$ for $i = 0, 1, 2, 3$ according to
(\ref{defTi}); i.e.,
\begin{align}
 T_i:  &=  \int_{ \mathcal{S}_i }
 e^{-n p_s |D_{{r_n}}(\hat{v_x})|} \, \textrm{d}\hat{v_x}. \nonumber
\end{align}
Then
 \begin{align}
\mathbb{P}[I_x]  &=\sum_{i=0}^{3}T_i. \label{mbtxIx}
  \end{align}

To compute $T_0$, we use $ D_{{r_n}}(\hat{v_x})  =  \pi {r_n}^2 $
for any position $ \hat{v_x} \in \mathcal{S}_0$, and $
|\mathcal{S}_0| = (1-2{r_n})^2 \leq 1$ to derive
\begin{align}
 T_0 &= \int_{ \mathcal{S}_0}
 e^{-p_s n |D_{{r_n}}(\hat{v_x})|}
\, \textrm{d}\hat{v_x} \leq e^{- \pi {r_n}^2 p_s  n}. \label{mbt0}
 \end{align}

 We present below an upper bound on $T_1$. In view of (\ref{go8t1}),
we further have
 \begin{align}
& -(H')^{-2}H'' e^{-p_s n H }  \, \textrm{d}g \nonumber
\\ &   \quad =  -(H')^{-3}H'' e^{-p_s n H
  } \, \textrm{d}H \nonumber
\\ & \quad = -(p_s n)^{-1} (H')^{-3}H''  \, \textrm{d}\big(-e^{-p_s n H
  }\big). \label{go8t1mcm}
 \end{align}
For $0 \leq g \leq \frac{{r_n}}{2}$, it holds from (\ref{h2}) and
(\ref{h3}) that
\begin{align}
 -\frac{H''}{(H')^3}& =  \frac{g}{4({r_n}^2-g^2)^2}
 \leq \frac{\frac{{r_n}}{2}}{4 \times
\left(\frac{3}{4}{r_n}^2\right)^2}
   = \frac{2}{9{r_n}^3}. \label{it2}
 \end{align}
By (\ref{go8t1mcm}) and (\ref{it2}), \f
\begin{align}
 -\int_{ 0}^{ \frac{{r_n}}{2}} (H')^{-2}H'' e^{-p_s n H }
 \textrm{d}g  &  \leq
  \frac{2}{9{r_n}^3p_s  n  }
    \int_{ 0}^{ \frac{{r_n}}{2}} \textrm{d}\big( - e^{-p_s n H
  }\big).  \label{it3mbst}
 \end{align}
 Applying (\ref{it3mbst}) to (\ref{go8t1}),
\begin{align}
&   \int_{ 0}^{ \frac{{r_n}}{2}}
 e^{-p_s n H(g)} \, \textrm{d}g   \leq \frac{ e^{-p_s n H(0) }}{p_s n H'(0)}%
+\frac{2 e^{-p_s n H(0)
  }}{9 {r_n}^3{p_s}^2 n^2 } . \label{rn2dg}
 \end{align}

Using (\ref{rn2dg}) in (\ref{go8}),%
%
\begin{align}
T_1  & \leq\frac{ 4  e^{-p_s n H(0) }}
 {p_s n H'(0)} +\frac{8 e^{-p_s n H(0)
  }}{9 {r_n}^3{p_s}^2 n^2 } . \label{T1nH0}
 \end{align}

From (\ref{h}) and (\ref{h2}), then $H(0)  = \pi {r_n}^2 / 2$ and'
$H'(0) = 2 r_n$. Using these and $\pi {r_n}^2 {p_s} n  = \Theta(\ln
n)$ from Lemma \ref{lems} in
(\ref{T1nH0}), 
 we derive
\begin{align}
T_1  & \leq 2 (r_n p_s n)^{-1} e^{-\pi {r_n}^2 p_s n / 2} \cdot
[1+o(1)] . \label{mbt1up}
 \end{align}

%

Now we assess $T_2$. For $\hat{v_x} \in \mathcal{S}_2$, when the
distance from $\hat{v_x}$ to the nearest edge of $\mathcal{S}$
equals $\frac{{r_n}}{2}$, the area $|D_{{r_n}}(\hat{v_x})|$ reaches
its minimum $c_ 0 \pi {r_n}^2$, where $c_ 0 : = \frac{2}{3}  +
\frac{\sqrt{3}}{4 \pi}$. Then with $|\mathcal{S}_2| = 2
{r_n}(1-2{r_n}) \leq 2 {r_n}$, it follows
\begin{align}
 T_2   & = \int_{ \mathcal{S}_2}
 e^{-p_s n |D_{{r_n}}(\hat{v_x})|}\,
  \textrm{d} \hat{v_x} \leq  2 {r_n}
   e^{- c_0 \pi {r_n}^2 p_s n} . \label{mbt2}
 \end{align}

To evaluate $T_3$, we apply $ { \pi {r_n}^2}/{4} \leq
D_{{r_n}}(\hat{v_x}) \leq   \pi {r_n}^2$ for any $\hat{v_x} \in
\mathcal{S}_3$, and $ |\mathcal{S}_3| = 4{r_n}^2$ to obtain
\begin{align}
 T_3 & = \int_{ \mathcal{S}_3}
 e^{-p_s n |D_{{r_n}}(\hat{v_x})|} \, \textrm{d}\hat{v_x} \leq
 4{r_n}^2  e^{- \pi {r_n}^2 p_s n /4} .\label{mbt3}
 \end{align}

 Substituting (\ref{mbt0}) (\ref{mbt1up}) (\ref{mbt2}) and
 (\ref{mbt3})
  into (\ref{mbtxIx}), we derive
\begin{align}
 \mathbb{P}[I_x]  & \leq e^{- \pi {r_n}^2 p_s  n} + 2 (r_n p_s n)^{-1} e^{-\pi {r_n}^2 p_s n /
 2} \cdot
[1+o(1)]
  \nonumber \\
   & \quad +  2 {r_n} e^{- c_0 \pi {r_n}^2 p_s
n} +  4{r_n}^2  e^{- \pi {r_n}^2 p_s n /4}.\label{pixup}
  \end{align}

  We discuss the following cases (i) and (ii), in which
  either $\frac{{K_n}^2}{P_n}  =  \omega \left(
\frac{1}{n^{1/3}\ln n} \right)$ or $\frac{{K_n}^2}{P_n}  =  O \left(
\frac{1}{n^{1/3}\ln n} \right)$ holds. We also let
  $\Delta$ denote
   $\pi {r_n}^2 p_s n$.


  (i) We consider $\frac{{K_n}^2}{P_n}  =  \omega \left(
\frac{1}{n^{1/3}\ln n} \right)$, which also yields $p_s =  \omega
\left( \frac{1}{n^{1/3}\ln n} \right)$ by (\ref{pssim}). Using
(\ref{mxrn2}) and (\ref{mxrn}) in (\ref{pixup}), we obtain
\begin{align}
 n\mathbb{P}[I_x] & \leq  p_s e^{-\delta_n} \ln \frac{n}{p_s}
\nonumber \\
   & \quad + 2 \pi^{\frac{1}{2}}
    \Delta^{ -\frac{1}{2}} \bigg(\ln \frac{n}{p_s}\bigg)^{\frac{1}{2}} e^{-\frac{\delta_n}{2}} \cdot
[1+o(1)]\nonumber \\
   & \quad + 2 \pi ^{-\frac{1}{2}}
   {p_s}^{c_0-\frac{1}{2}}n^{\frac{1}{2}-c_0}
    \Delta^{ \frac{1}{2} } \bigg(\ln \frac{n}{p_s}\bigg)^{c_0}
  e^{- c_0 \delta_n}
\nonumber \\
   & \quad +  4 \pi ^{-1}
     {p_s}^{- \frac{3}{4}}  n^{- \frac{ 1}{4}}  \Delta
 e^{-\frac{\delta_n}{4}}
 \bigg(\ln \frac{n}{p_s}\bigg)^{ \frac{1}{4}} . \nonumber
 \end{align}
From $p_s =  \omega \left( \frac{1}{n^{1/3}\ln n} \right)$, $\Delta
= \Theta(\ln n)$ and $ \ln \frac{n}{p_s} = \Theta(\ln n)$ by Lemma
\ref{lems}, with $ |\alpha_n|= O(\ln \ln n)$ producing $ |\delta_n|=
O(\ln \ln n)$, we have
\begin{align}
  & n\mathbb{P}[I_x] = o(n^{\epsilon})\textrm{ for any constant }\epsilon>0. \nonumber
 \end{align}

(ii) We consider $\frac{{K_n}^2}{P_n}  =  O \left(
\frac{1}{n^{1/3}\ln n} \right)$, which also yields $p_s =  O \left(
\frac{1}{n^{1/3}\ln n} \right)$ by (\ref{pssim}).  We obtain
\begin{align}
r_n & = \pi^{-\frac{1}{2}}{p_s}^{-\frac{1}{2}}n^{-\frac{1}{2}}
\Delta^{\frac{1}{2}}, \la{mxrncase2}
\end{align}
and from (\ref{caseiirp}),
\begin{align}
e^{-\pi {r_n}^2 p_s n} & = e^{-4\ln \frac{1}{p_s} + 4\ln \ln
\frac{1}{p_s} - \delta_{n}} = {p_s}^4 \bigg(\ln
\frac{1}{p_s}\bigg)^4 e^{- \delta_{n}}. \la{mxrn2case2}
\end{align}
Applying (\ref{mxrncase2}) and (\ref{mxrn2case2}) to (\ref{pixup}),
we derive
\begin{align}
 n\mathbb{P}[I_x] & \leq  {p_s}^4 \bigg(\ln
\frac{1}{p_s}\bigg)^4 e^{- \delta_{n}}
 \nonumber \\
   & \quad +  2 \pi^{\frac{1}{2}}{p_s}^{\frac{3}{2}}n^{\frac{1}{2}}
    \Delta^{ -\frac{1}{2}} \bigg(\ln \frac{n}{p_s}\bigg)^{2} e^{-\frac{\delta_n}{2}}\nonumber \\
   & \quad + 2 \pi ^{-\frac{1}{2}}
   {p_s}^{4c_0-\frac{1}{2}}n^{\frac{1}{2}}
    \Delta^{ \frac{1}{2} } \bigg(\ln \frac{1}{p_s}\bigg)^{4c_0}
  e^{- c_0 \delta_n}  \nonumber \\
   & \quad +4 \pi ^{-1}  \Delta
 e^{-\frac{\delta_n}{4}} \ln \frac{1}{p_s}
.\nonumber
 \end{align}
From $\Delta = \Theta(\ln n)$ and $ \ln \frac{1}{p_s} = \Theta(\ln
n)$, $p_s =  O \left( \frac{1}{n^{1/3}\ln n} \right)$ and $
\frac{1}{3}(4c_0-\frac{1}{2})>\frac{1}{2}$, with $ |\alpha_n|= O(\ln
\ln n)$ producing $ |\delta_n|= O(\ln \ln n)$, we have
\begin{align}
  & n\mathbb{P}[I_x] = o(n^{\epsilon})\textrm{ for any constant }\epsilon>0. \nonumber
 \end{align}

%

\subsection{The Proof of Lemma \ref{dePoissonization}}

We will establish Lemma \ref{dePoissonization} using the standard
de-Poissonization technique
\cite{citeulike:505396,penrose,WanCoverage}. Let $M$ be the number
of nodes in graph $G_{\textrm{Poisson}}
  (m, \theta_n, \mathcal{A})$.
Clearly, $M$ follows a Poisson distribution with mean $m$. From
Chebyshev's inequality, for any positive $t$,
\begin{align}
\mathbb{P}[|M - m| \geq t\sqrt{m}] & \leq t^{-2}.\label{eqmm}
\end{align}
Without loss of generosity, we regard $n-n^{\frac{1}{2}+c_0}$ as an
integer. With $t:=\frac{n-m}{\sqrt{m}}$, substituting $m =
 n-n^{\frac{1}{2}+c_0} $ into (\ref{eqmm}),
\begin{align}
 & \mathbb{P}\big[ M \leq n-2n^{\frac{1}{2}+c_0}\textrm{ or }M \geq
n\big] \leq \frac{n-n^{\frac{1}{2}+c_0}}{n^{1+2c_0}} = o(1).
\label{eqmmjyzq}
\end{align}
Hence, $ n-2n^{\frac{1}{2}+c_0} < M < n$ holds almost surely.

When $M < n$, we construct a coupling $\mathcal {C}$ between graphs
$G
  (n, \theta_n, \mathcal{A})$ and $G_{\textrm{Poisson}}
  (m, \theta_n, \mathcal{A})$, by letting
 the former be the result of adding to
 the latter graph $ (n-M) $ nodes uniformly
 distributed on $\mathcal{A}$. Then $\mathcal {V}_P$ denoting the node
  set of $G_{\textrm{Poisson}}
  (m, \theta_n, \mathcal{A})$ is a subset of $\mathcal {V}$ being the node
  set of $G
  (n, \theta_n, \mathcal{A})$. In addition, it is straightforward to
  see that the edge set of $G_{\textrm{Poisson}}
  (m, \theta_n, \mathcal{A})$ is also a
  subset of that of $G
  (n, \theta_n, \mathcal{A})$. Then under coupling $\mathcal {C}$,
graph $G_{\textrm{Poisson}}
  (m, \theta_n, \mathcal{A})$
is a subgraph of $G
  (n, \theta_n, \mathcal{A})$.

We denote by $D_X$ (resp., $D_P$) the set of isolated
   nodes in $G
  (n, \theta_n, \mathcal{A})$ (resp., $G_{\textrm{Poisson}}
  (m, \theta_n, \mathcal{A})$). To establish Lemma
  \ref{dePoissonization}, we prove
\begin{align}
| \mathbb{P}[D_X \neq \emptyset] - \mathbb{P} [D_P \neq \emptyset] |
&   = o(1),\nonumber
\end{align}
which follows once we demonstrate $\mathbb{P}[D_X\neq D_P] = o(1)$
in view of
\begin{align}
\lefteqn{ \big| \mathbb{P}[D_X \neq \emptyset] - \mathbb{P} [D_P
\neq \emptyset] \big|} \nonumber
\\  & = \big|\mathbb{P}[(D_X \neq \emptyset) \hspace{-1pt}\bcap\hspace{-1pt} (D_P
= \emptyset) ] - \mathbb{P}[(D_P \neq \emptyset) \hspace{-1pt}\bcap
\hspace{-1pt}(D_X = \emptyset) ]\big| \nonumber
\\  & \leq \mathbb{P}[D_X\neq D_P]. \nonumber
\end{align}
It is straightforward to see
\begin{align}
 & \mathbb{P}[D_X\neq D_P] \nonumber
\\  & \leq \mathbb{P}\big[ (D_P \setminus D_X \neq \emptyset) \bcap (n-2n^{\frac{1}{2}+c_0} < M < n)\big]
 \nonumber
\\  & \quad +
\mathbb{P}\big[ (D_X \setminus D_P \neq \emptyset) \bcap
(n-2n^{\frac{1}{2}+c_0} < M < n)\big]\nonumber
\\  & \quad +
\mathbb{P}\big[ M \leq n-2n^{\frac{1}{2}+c_0}\textrm{ or }M \geq
n\big] ,\nonumber
\end{align}
then given (\ref{eqmmjyzq}), we will prove $\mathbb{P}[D_X\neq D_P]
= o(1)$ and thus establish Lemma \ref{dePoissonization} once showing
\begin{align}
\mathbb{P}\big[ (D_P \setminus D_X \neq \emptyset) \bcap
(n-2n^{\frac{1}{2}+c_0} < M < n)\big] &   = o(1), \label{eqnn31}
\end{align}
and
\begin{align}
\mathbb{P}\big[ (D_X \setminus D_P \neq \emptyset) \bcap
(n-2n^{\frac{1}{2}+c_0} < M < n)\big] &   = o(1). \label{eqnn32}
\end{align}

In proving (\ref{eqnn31}) and (\ref{eqnn32}), with
$n-2n^{\frac{1}{2}+c_0} < M < n$, we consider the coupling $\mathcal
{C}$ under which $G_{\textrm{Poisson}}
  (m, \theta_n, \mathcal{A})$
is a subgraph of $G
  (n, \theta_n, \mathcal{A})$.

\subsubsection{The Proof of (\ref{eqnn31})}

 Event ($D_P \setminus D_X \neq \emptyset$) happens if and only
if there exists at least one node $v_i$ such that $v_i \in D_P $ and
$v_i \notin D_X $; i.e., $v_i$ is isolated in $G_{\textrm{Poisson}}
(m, \theta_n, \mathcal{A})$ but is not isolated in $G
  (n, \theta_n, \mathcal{A})$ since $v_i \in \mathcal {V}$ by $v_i \in D_P $. Then there exists at least one node $v'$ in
$\mathcal {V}_{\overline{P}} := \mathcal {V} \setminus \mathcal
{V}_P$ such that $v'$ and $v_i$ are neighbors in $G   (n, \theta_n,
\mathcal{A})$. Due to $\mathcal {V}_{\overline{P}} = n - M <
2n^{\frac{1}{2}+c_0}$, considering that $p_e$ is the edge
probability in $G   (n, \theta_n, \mathcal{A})$, then with $L$
denoting the number of isolated nodes in $G_{\textrm{Poisson}}
  (m, \theta_n, \mathcal{A})$, it follows via a union bound that
\begin{align}
& \mathbb{P}\big[ \big(D_P \setminus D_X \neq \emptyset\big) \bcap
\big(n-2n^{\frac{1}{2}+c_0} < M < n\big)\big] \nonumber
\\ & \quad \leq L \cdot 2n^{\frac{1}{2}+c_0} \cdot p_e. \label{eqnn22zkjalfjs}
\end{align}

For $\mathcal{A} = \mathcal{T}$, we will prove that under conditions
(\ref{thm:t_Kn}) and (\ref{thm:t:rnKnPn}) with $|\alpha_n| = O(\ln
\ln n)$; i.e., with some $\mu_n = \omega(1)$ and constant $c_1$,
\begin{align}
\max\bigg\{ \frac{\ln n}{\ln \ln n}, \hspace{1.5pt} \mu_n \cdot
\sqrt{\frac{P_n \ln n}{n}}
 \bigg\} &  \leq   K_n  \leq   c_1 \sqrt{\frac{P_n}{\ln n}} ,\nonumber
\end{align}
for all $n$ sufficiently large, and
\begin{align}
 \pi {r_n}^2 \cdot \frac{{K_n}^2}{P_n} & = \frac{\ln n +
 \alpha_{n}}{n},\nonumber
\end{align}
then there exist some $\widetilde{\mu}_n = \omega(1)$ and constant
$\widetilde{c}_1$ such that
\begin{align}
\max\bigg\{ \frac{\ln m}{\ln \ln m}, \sqrt{\frac{P_n \ln m}{m}}
\cdot \widetilde{\mu}_n \bigg\} & \leq K_n \leq  \widetilde{c}_1
\sqrt{\frac{P_n}{\ln m}} \label{thm:t_Kn_pm}
\end{align}
for all $m$ sufficiently large (i.e., for all $n$ sufficiently
large) and
\begin{align}
 \pi {r_n}^2 \cdot \frac{{K_n}^2}{P_n} & = \frac{\ln m +
 \alpha_{n} \pm o(1)}{m}, \label{thm:t:rnKnPn_pm}
\end{align}
in order to apply Lemma \ref{Poissonization1} to
$G_{\textrm{Poisson}}
  (m, \theta_n, \mathcal{T})$.

We establish (\ref{thm:t_Kn_pm}) as follows. First, with $m < n$, \f
$\frac{\ln m}{\ln \ln m} < \frac{\ln n}{\ln \ln n}$ for all $n$
sufficiently large since function $f(x) = \frac{\ln x}{\ln \ln x}$
is monotone increasing with $x$ for $x>e^e$ by $f'(x) = \frac{\ln
\ln x-1}{x(\ln \ln x)^2} > 0$. Thus, under $K_n \geq \frac{\ln
n}{\ln \ln n}$, it holds that $K_n \geq \frac{\ln m}{\ln \ln m}$.
Second, with $m = n-n^{\frac{1}{2}+c_0}$ and $0 < c_0 <
\frac{1}{2}$, then $\frac{\ln m}{m} \sim \frac{\ln n}{n}$, which
along with $K_n \geq \sqrt{\frac{P_n \ln n}{n}} \cdot \mu_n$ and
$\mu_n = \omega(1)$ leads to that there exists some
$\widetilde{\mu}_n = \omega(1)$ such that $K_n \geq \sqrt{\frac{P_n
\ln m}{m}} \cdot \widetilde{\mu}_n$. Third, with $m \sim n$, for all
$n$ sufficiently large, we obtain from $K_n  \leq   c_1
\sqrt{\frac{P_n}{\ln n}}$ that $K_n \leq  \widetilde{c}_1
\sqrt{\frac{P_n}{\ln m}}$. Summarizing the three points above,
(\ref{thm:t_Kn_pm}) holds for all $m$ sufficiently large.

We demonstrate (\ref{thm:t:rnKnPn_pm}) in view of 
%
%
\begin{align}
 & m \cdot \pi {r_n}^2 \cdot \frac{{K_n}^2}{P_n}- (\ln m +
 \alpha_{n})  \nonumber\\ & = m \cdot \frac{\ln n +
 \alpha_{n}}{n} - (\ln m +
 \alpha_{n}) \nonumber\\ & = \big(1-n^{c_0-\frac{1}{2}}\big)(\ln n +
 \alpha_{n}) - \ln n -
  \ln \big(1-n^{c_0-\frac{1}{2}}\big) - \alpha_{n} \nonumber\\ & =
   \pm o(1). \nonumber
\end{align}
Then with (\ref{thm:t_Kn_pm}) and (\ref{thm:t:rnKnPn_pm}), we use
Lemma \ref{Poissonization1} to derive
\begin{align}
  & L = o(m^{\epsilon})\textrm{ for any constant
  }\epsilon>0. \label{tL}
 \end{align}

For $\mathcal{A} = \mathcal{S}$, we will prove that under conditions
(\ref{thm:s_Kn}) and (\ref{thm:s:rnKnPn}) with $|\alpha_n| = O(\ln
\ln n)$; i.e., with constants $c_2
>0,$ $0<c_3<1$, $c_4>0$ and $\nu_n = o(1)$,
\begin{align}
 c_2 \sqrt{\frac{P_n \ln n}{n^{c_3}}} & \leq K_n \leq
\min\Bigg\{\nu_n \cdot  \sqrt{\frac{P_n}{\ln n}},
\hspace{1.5pt}{\frac{ c_4 P_n}{n\ln n}}\hspace{2pt}\Bigg\} \nonumber
\end{align}
for all $n$ sufficiently large, the condition that
$\frac{{K_n}^2}{P_n} \cdot n^{1/3}\ln n$ \emph{either} is bounded
for all $n$ \emph{or} converges to $\infty$ as $n \to \infty$, and
\begin{align}
\hspace{-117pt} \pi {r_n}^2 \cdot  \textrm{\fontsize{9.5}{11}
\selectfont $\frac{{K_n}^2}{P_n}$} & \nonumber
\end{align}\vspace{-10pt}
\begin{align}
&\hspace{-4pt}=\begin{cases} \hspace{-4pt}\textrm{\fontsize{13}{15}
\selectfont $\frac{\ln \frac{n P_{n}}{{K_{n}}^2}
\hspace{2pt}-\hspace{2pt} \ln \ln \frac{n P_{n}}{{K_{n}}^2}
\hspace{2pt}+\hspace{2pt} \alpha_{n} }{n}$} ,
&\hspace{-6pt}\textrm{for~} \textrm{\large \selectfont
$\frac{{K_n}^2}{P_n}$} = \omega
\textrm{\large \selectfont $\left( \frac{1}{n^{1/3}\ln n} \right)$}, \\
\vspace{-7pt} \\
 \hspace{-4pt}\textrm{\fontsize{13}{15} \selectfont $\frac{ 4\ln \frac{P_n}{{K_n}^2} \hspace{2pt}
 -\hspace{2pt} 4\ln \ln \frac{P_n}{{K_n}^2} \hspace{2pt}+\hspace{2pt}
\alpha_{n}  }{n}$} , &\hspace{-6pt}\textrm{for~} \textrm{\large
\selectfont $\frac{{K_{n}}^2}{P_{n}}$}  = O \textrm{\large
\selectfont $\left( \frac{1}{n^{1/3}\ln n} \right)$};\end{cases}
\nonumber
\end{align}
then there exist some constants $\widetilde{c}_2
>0$, $0<\widetilde{c}_3<1$, $\widetilde{c}_4>0$ and $\widetilde{\nu}_n = o(1)$,
 such that
\begin{align}
 \widetilde{c}_2 \sqrt{\frac{P_n \ln m}{m^{\widetilde{c}_3}}} &
  \leq K_n \leq \min\Bigg\{\widetilde{\nu}_n \cdot  \sqrt{\frac{P_n}{\ln m}},
\hspace{1.5pt}{\frac{ \widetilde{c}_4 P_n}{m\ln
m}}\hspace{2pt}\Bigg\} \label{thm:s_Kn_pm}
\end{align}
for all $m$ sufficiently large (i.e., for all $n$ sufficiently
large); and $\frac{{K_n}^2}{P_n} \cdot m^{1/3}\ln m$ \emph{either}
is bounded for all $m$ \emph{or} converges to $\infty$ as $m \to
\infty$ (i.e., as $n \to \infty$); and $\widetilde{\alpha}_n =
\alpha_n \pm O(1)$ with sequence $\widetilde{\alpha}_n$ for all $n$
defined through
\begin{align}
\hspace{-117pt} \pi {r_n}^2 \cdot \frac{{K_n}^2}{P_n}  & \nonumber
\end{align}\vspace{-7pt}
\begin{align}
&\hspace{-4pt}=\begin{cases} \hspace{-4pt}\textrm{\fontsize{13}{15}
\selectfont $\frac{\ln \frac{m P_{n}}{{K_{n}}^2}
\hspace{2pt}-\hspace{2pt} \ln \ln \frac{m P_{n}}{{K_{n}}^2}
\hspace{2pt}+\hspace{2pt} \widetilde{\alpha}_{n} }{m}$} ,
\textrm{~for~} \textrm{\large \selectfont $\frac{{K_n}^2}{P_n}$} =
\omega
\textrm{\large \selectfont $\left( \frac{1}{m^{1/3}\ln m} \right)$}, \\
\vspace{-7pt} \\
 \hspace{-4pt}\textrm{\fontsize{13}{15} \selectfont $\frac{ 4\ln \frac{P_n}{{K_n}^2} \hspace{2pt}
 -\hspace{2pt} 4\ln \ln \frac{P_n}{{K_n}^2} \hspace{2pt}+\hspace{2pt}
\widetilde{\alpha}_{n}  }{m}$} , \textrm{~for~} \textrm{\large
\selectfont $\frac{{K_{n}}^2}{P_{n}}$}  = O \textrm{\large
\selectfont $\left( \frac{1}{m^{1/3}\ln m} \right)$},\end{cases}
\label{thm:s:rnKnPn_pm}
\end{align}
in order to apply Lemma \ref{Poissonization2} to
$G_{\textrm{Poisson}}
  (m, \theta_n, \mathcal{S})$.

We establish (\ref{thm:s_Kn_pm}) as follows. First, due to $K_n \geq
c_2 \sqrt{\frac{P_n \ln n}{n^{c_3}}}$ and $\frac{\ln m}{m^{c_3}}\sim
\frac{\ln n}{n^{c_3}}$, then with $\widetilde{c}_3 = c_3$, there
exits some $\widetilde{c}_2
>0$ such that $K_n \geq
\widetilde{c}_2 \sqrt{\frac{P_n \ln m}{m^{\widetilde{c}_3}}}$.
Second, from $\ln m \sim \ln n$ and $K_n \leq \nu_n \cdot
\sqrt{\frac{P_n}{\ln n}}$, there exists some $\widetilde{\nu}_n =
o(1)$ such that $K_n \leq \widetilde{\nu}_n \cdot
\sqrt{\frac{P_n}{\ln m}}$. Third, from $m \ln m \sim n\ln n$ and
$K_n \leq \frac{{c}_4 P_n}{n\ln n}$, there exists some
$\widetilde{c}_4>0$ such that $K_n \leq \frac{ \widetilde{c}_4
P_n}{m\ln m}$. Summarizing the three points above,
(\ref{thm:s_Kn_pm}) holds for all $m$ sufficiently large.

We demonstrate (\ref{thm:s:rnKnPn_pm}) below. Clearly, it holds that
$m^{1/3}\ln m \sim n^{1/3}\ln n$. This implies that condition
$\frac{{K_{n}}^2}{P_{n}} = \omega\left( \frac{1}{m^{1/3}\ln m}
\right)$ (resp., $\frac{{K_{n}}^2}{P_{n}} = O\left(
\frac{1}{m^{1/3}\ln m} \right)$) is equivalent to
$\frac{{K_{n}}^2}{P_{n}} = \omega\left( \frac{1}{n^{1/3}\ln n}
\right)$ (resp., $\frac{{K_{n}}^2}{P_{n}} = O\left(
\frac{1}{n^{1/3}\ln n} \right)$).

On the one hand, for $\frac{{K_{n}}^2}{P_{n}} = \omega\left(
\frac{1}{m^{1/3}\ln m} \right)$ equivalent to
$\frac{{K_{n}}^2}{P_{n}} = \omega\left( \frac{1}{n^{1/3}\ln n}
\right)$, we obtain
\begin{align}
 &  \widetilde{\alpha}_{n} \nonumber \\  & = m \cdot  \pi {r_n}^2 \cdot \frac{{K_n}^2}{P_n}
- \bigg(\ln \frac{m P_{n}}{{K_{n}}^2} - \ln \ln \frac{m
P_{n}}{{K_{n}}^2}\bigg)\nonumber \\  & =  m \cdot \frac{\ln \frac{n
P_{n}}{{K_{n}}^2} \hspace{-1pt} - \hspace{-1pt} \ln \ln \frac{n
P_{n}}{{K_{n}}^2} \hspace{-1pt}+ \hspace{-1pt} \alpha_{n} }{n}
\hspace{-1pt} - \hspace{-1pt} \bigg( \hspace{-2pt}\ln \frac{m
P_{n}}{{K_{n}}^2} \hspace{-1pt} - \hspace{-1pt} \ln \ln \frac{m
P_{n}}{{K_{n}}^2}\bigg)\nonumber \\  & = \alpha_n \pm O(1) .
\nonumber
\end{align}

On the other hand, for $\frac{{K_{n}}^2}{P_{n}} = O\left(
\frac{1}{m^{1/3}\ln m} \right)$ equivalent to
$\frac{{K_{n}}^2}{P_{n}} = O\left( \frac{1}{n^{1/3}\ln n} \right)$,
we obtain
\begin{align}
  &\widetilde{\alpha}_{n} \nonumber \\
    & \hspace{-2pt}= \hspace{-2pt} m \cdot  \pi {r_n}^2 \cdot \frac{{K_n}^2}{P_n}
- \bigg(4\ln \frac{P_n}{{K_n}^2}
 -  4\ln \ln \frac{P_n}{{K_n}^2}\bigg)\nonumber \\  & \hspace{-2pt} = \hspace{-2pt}  m \hspace{-2pt}
 \cdot \hspace{-2pt}
 \frac{4\ln \frac{P_n}{{K_n}^2} \hspace{-2pt}
 - \hspace{-2pt} 4\ln \ln \frac{P_n}{{K_n}^2} \hspace{-2pt}+ \hspace{-2pt} \alpha_{n} }{n}
\hspace{-2pt} - \hspace{-2pt} \bigg( \hspace{-2pt}4\ln
\frac{P_n}{{K_n}^2} \hspace{-2pt} - \hspace{-2pt} 4\ln \ln
\frac{P_n}{{K_n}^2} \hspace{-2pt}\bigg)\nonumber \\  & \hspace{-2pt}
= \hspace{-2pt} \alpha_n \pm O(1) . \nonumber
\end{align}
Then with (\ref{thm:s_Kn_pm}) and (\ref{thm:s:rnKnPn_pm}), we use
Lemma \ref{Poissonization2} to derive
\begin{align}
  & L = o(m^{\epsilon})\textrm{ for any constant
  }\epsilon>0. \nonumber
 \end{align}

To summarize, for either $\mathcal{A} = \mathcal{T}$ or $\mathcal{A}
= \mathcal{S}$, it always holds that $L = o(m^{\epsilon}) =
o(n^{\epsilon})$ for any constant $\epsilon>0$, where $L$ is the
number of isolated nodes in $G_{\textrm{Poisson}}
  (m, \theta_n, \mathcal{A})$. For $\mathcal{A} = \mathcal{T}$, we have $p_e =  \pi {r_n}^2 \cdot
p_s $. For $\mathcal{A} = \mathcal{S}$, we obtain $p_e \leq \pi
{r_n}^2 \cdot p_s $. Also, we have $\pi {r_n}^2 \cdot p_s =
\Theta\big(\frac{\ln n}{n}\big)$. Then from (\ref{eqnn22zkjalfjs}),
with $\epsilon$ set as $0< \epsilon < \frac{1}{2}-c_0$,  \f
\begin{align}
& \mathbb{P}\big[ \big(D_P \setminus D_X \neq \emptyset\big) \bcap
\big(n-2n^{\frac{1}{2}+c_0} < M < n\big)\big] \nonumber
\\ & \quad \leq o(n^{\epsilon}) \cdot 2n^{\frac{1}{2}+c_0}
\cdot \Theta\bigg(\frac{\ln n}{n}\bigg) = o(1). \nonumber
\end{align}

\subsubsection{The Proof of (\ref{eqnn32})}

Event ($D_X \setminus D_P \neq \emptyset$) occurs if and only if
there exists at least one node $v_j$ such that $v_j \in D_X $ and
$v_j \notin D_P $. With $v_j \in D_X $, then $v_j$ is isolated in $G
  (n, \theta_n, \mathcal{A})$, which along with $v_j \notin D_P $
  leads to $v_j  \notin \mathcal {V}_{ {P}}$ and
  $v_j  \in \mathcal
{V}_{\overline{P}}$ (i.e., $v_j$ is not a node in
  graph $G_{\textrm{Poisson}} (m, \theta_n,
\mathcal{A})$). This can be seen by contradiction. Supposing $v_j
\notin \mathcal {V}_{ {P}}$, since $v_j$ is isolated in $G
  (n, \theta_n, \mathcal{A})$, then $v_j$ is also isolated in $G_{\textrm{Poisson}} (m, \theta_n,
\mathcal{A})$, contradicting $v_j \notin D_P $. Then with $q$
denoting the probability that a node is isolated in $G
  (n, \theta_n, \mathcal{A})$, it follows via a union bound that
\begin{align}
& \mathbb{P}\big[ \big(D_X \setminus D_P \neq \emptyset\big) \bcap
\big(n-2n^{\frac{1}{2}+c_0} < M < n\big)\big] \nonumber
\\ & \quad \leq 2n^{\frac{1}{2}+c_0} \cdot q. \nonumber
\end{align}
Due to $q = \Theta\big(\frac{\ln n}{n}\big)$ and $c_0<\frac{1}{2}$,
then
\begin{align}
& \mathbb{P}\big[ \big(D_X \setminus D_P \neq \emptyset\big) \bcap
\big(n-2n^{\frac{1}{2}+c_0} < M < n\big)\big] = o(1). \nonumber
\end{align}

\subsection{The Proof of Lemma \ref{erg_rgg_conn:t}}

By \cite[Theorem 2.3]{Penrose2013}, if there exists some $ \beta \in
[0, \infty]$ such that
\begin{align}
 \lim_{n \to \infty} n  \int_{ \mathcal{T} }
 e^{-n p_s |D_{{r_n}}(\hat{v_x})|} \, \textrm{d}\hat{v_x}  & = \beta
  , \label{bt1}
\end{align}
then
\begin{align}
 \lim_{n \to \infty} \mathbb{P}\left[\hspace{-1pt}
  \begin{array}{l} G_{ER}(n,p_n) \bcap G_{RGG}(n, r_n, \mathcal{T}\hspace{1pt}) \\
 \textrm{is connected.}
\end{array} \hspace{-1pt}\right] & = e^{-\beta}. \label{bt2}
\end{align}
Given $|D_{{r_n}}(\hat{v_x})| = \pi {r_n}^2$ for any $\hat{v_x} \in
\mathcal{T}$, we obtain
\begin{align}
n  \int_{ \mathcal{T} }
 e^{-n p_n |D_{{r_n}}(\hat{v_x})|} \, \textrm{d}\hat{v_x} &=
 n \cdot e^{-\pi {r_n}^2 p_n n} \cdot |\mathcal{T}| \nonumber \\
 & =
n e^{-\pi {r_n}^2
 p_n n}. \label{prIxenew}
\end{align}
Applying condition $\pi {r_n}^2 p_n n = \ln n +
 \nu_n$ to (\ref{prIxenew}), we have
\begin{align}
 \lim_{n \to \infty} n  \int_{ \mathcal{T} }
 e^{-n p_n |D_{{r_n}}(\hat{v_x})|} \, \textrm{d}\hat{v_x}  & =  \lim_{n \to \infty} e^{-\nu_n}
  \nonumber \\ & =
\begin{cases} \infty, \hspace{-2pt} &\textrm{if $\lim\limits_{n \to \infty}{\nu_n}
=-\infty$}, \\  0, \hspace{-2pt} &\textrm{if $\lim\limits_{n \to
\infty}{\nu_n} =\infty$.}\end{cases} \label{bt3}
\end{align}
In view of (\ref{bt1}) (\ref{bt2}) and (\ref{bt3}), \f
\begin{align}
&  \lim_{n \to \infty} \mathbb{P}\left[\hspace{-1pt}
  \begin{array}{l} G_{ER}(n,p_n) \bcap G_{RGG}(n, r_n, \mathcal{T}\hspace{1pt}) \\
 \textrm{is connected.}
\end{array} \hspace{-1pt}\right]  \nonumber \\ & \quad =
\begin{cases} e^{-\infty}, &\textrm{if $\lim\limits_{n \to \infty}{\nu_n}
=-\infty$}, \\  e^{0}, &\textrm{if $\lim\limits_{n \to
\infty}{\nu_n} =\infty$,}\end{cases} \nonumber \\ & \quad =
\begin{cases} 0, & \hspace{14pt}\textrm{if $\lim\limits_{n \to \infty}{\nu_n}
=-\infty$}, \\  1, &\hspace{14pt}\textrm{if $\lim\limits_{n \to
\infty}{\nu_n} =\infty$.}\end{cases}
\end{align}
Hence, Lemma \ref{erg_rgg_conn:t} is proved.

%

\subsection{The Proof of Lemma \ref{cp_rig_er}}

We first explain the idea of coupling between random graphs. As used
by Rybarczyk \cite{zz}, a coupling of two random graphs $G_1$ and
$G_2$ means a probability space on which random graphs $G_1'$ and
$G_2'$ are defined such that $G_1'$ and $G_2'$ have the same
distributions as $G_1$ and $G_2$, respectively. We denote the
coupling by $(G_1, G_2, G_1', G_2')$.

Following Rybarczyk's notation \cite{zz}, we write
\begin{align}
  G_1 \preceq_{1-o(1)} G_2
\label{g1g2coupling}
\end{align}
if there exists a coupling $(G_1, G_2, G_1', G_2')$, such that under
the coupling $G_1'$ is a subgraph of $G_2'$ with probability
$1-o(1)$. 

We then describe a graph model called random intersection graph,
which has been extensively studied in the literature. A random
intersection graph denoted by $G_{RIG}(n,P_n,p_n)$ is defined on $n$
nodes as follows. There exist a key pool of size $P_n$; and each key
in the pool is added to each sensor with probability $p_n$. 

In view of \cite[Lemma 4]{Rybarczyk},  if
\begin{align}
p_n P_n & = \omega\left( \ln n \right) \label{pnPn}
\end{align}
and for all $n$ sufficiently large\footnote{The term ``for all $n$
sufficiently large'' means ``for any $n \geq N$, where $N $ is
selected appropriately''.},
\begin{align}
K_n  & \geq p_n P_n + \sqrt{3(p_n P_n + \ln n) \ln n} ,
\label{Kngeq}
\end{align}
then
\begin{align}
G_{RIG}(n, p_n, P_n) & \preceq_{1-o(1)} G_{RKG}(n, K_n, P_n ) .
\label{cp_rig_rkg}
\end{align}

By \cite[Lemma 3]{zz}, if
\begin{align}
p_n &=o\left(1/n\right), \label{cp_c3}
\end{align}
and for all $n$ sufficiently large,
\begin{align}
{p_n}^2 P_n  & <   1, \label{cp_c4}
\end{align}
 with $s_n$ defined
through
 \begin{align}
s_n &: = {p_n}^2 P_n \cdot \left(1 - n{p_n} + 2 {p_n} -
\frac{{p_n}^2 P_n}{2} \right), \label{cp_c5}
\end{align}
then
\begin{align}
G_{ER}(n, s_n) & \preceq_{1-o(1)} G_{RIG}(n, p_n, P_n) .
\label{cp_erg_rig}
\end{align}

By \cite{2013arXiv1301.0466R}, the relation of
``$\preceq_{1-o(1)}$'' is transitive. In other words, for any three
graphs $G_a$, $G_b$ and $G_c$, if $G_a \preceq_{1-o(1)} G_b$ and
$G_b \preceq_{1-o(1)} G_c$, then $G_a \preceq_{1-o(1)} G_c$. Then
given (\ref{cp_rig_rkg}) and (\ref{cp_erg_rig}), we obtain that
under (\ref{pnPn}) (\ref{Kngeq}) (\ref{cp_c3}) (\ref{cp_c4}) and
(\ref{cp_c5}), it follows that
\begin{align}
G_{ER}(n, s_n) & \preceq_{1-o(1)} G_{RKG}(n, K_n, P_n ) .
\label{cp_erg_rkg_ieq}
\end{align}
By \cite{2013arXiv1301.0466R}, from (\ref{cp_erg_rkg_ieq}), it
further holds that
\begin{align}
& G_{ER}(n, s_n) \bcap G_{RGG}(n, r_n, \mathcal{A})  \nonumber
\\  & \hspace{2pt}
\preceq_{1-o(1)} \big[ G_{RKG}(n, K_n, P_n ) \bcap G_{RGG}(n, r_n,
\mathcal{A}) \big]. \label{cp_erg_rkg_rgg}
\end{align}
By \cite{2013arXiv1301.0466R}, from (\ref{cp_erg_rkg_rgg}), it is
easy to see that for any monotone increasing graph property
$\mathscr{P}$,
\begin{align}
& \mathbb{P}[ \hspace{2pt}G_{RKG}(n, K_n, P_n) \bcap G_{RGG}(n, r_n,
\mathcal{A}) \textrm{ has }\mathscr{P}\hspace{2pt}] \nonumber
\\ & \quad \geq
 \mathbb{P}[ \hspace{2pt}G_{ER}(n, q_n) \bcap G_{RGG}(n, r_n,
\mathcal{A}) \textrm{ has }\mathscr{P}\hspace{2pt}] - o(1).
 \label{cp_erg_rkg_rgg_ineq}
 \end{align}
In view of (\ref{cp_erg_rkg_rgg_ineq}), the proof of Lemma
\ref{cp_rig_er} is completed with $q_n$ set as $s_n$ if we show that
given some
 appropriately
selected $p_n$ and the conditions in Lemma \ref{cp_rig_er}, then
(\ref{pnPn}) (\ref{Kngeq}) (\ref{cp_c3}) (\ref{cp_c4}) and
\begin{align}
s_n  & = \frac{{K_n}^2}{P_n} \cdot \left[1 -
 O \left(\frac{1}{ \ln n}\right)\right] \label{sn_prf}
\end{align}
with $s_n$ defined in (\ref{cp_c5}) all follow.

We will do so by setting $p_n$ via
\begin{align}
p_n  & = \frac{K_n}{P_n}
 \bigg(1 - \sqrt{\frac{3\ln
n}{K_n }}\hspace{2pt}\bigg) . \label{pn}
\end{align}
To begin with, from (\ref{pn}) and condition $K_n = \Omega\left((\ln
n)^3\right)$, it is clear that
\begin{align}
p_n  & = \frac{K_n}{P_n} \cdot [1-o(1)],\label{pn7}
\end{align}
which along with $K_n = \Omega\left((\ln n)^3\right)$ further leads
to
\begin{align}
p_n P_n   & = K_n   \cdot [1-o(1)]  = \Omega\left((\ln n)^3\right)
  = \omega\left( \ln n \right). \nonumber
\end{align}
Given (\ref{pn}) and condition $K_n = \Omega\left((\ln n)^3\right)$,
we obtain (\ref{Kngeq}) in that for all $n$ sufficiently large,
\begin{align}
&  K_n - \left[ p_n P_n + \sqrt{3(p_n P_n + \ln n) \ln n}
\hspace{1.5pt}\right] \nonumber \\ & = K_n \sqrt{\frac{3\ln n}{K_n
}} - \sqrt{3\left[ K_n \left(1 - \sqrt{\frac{3\ln n}{K_n
}}\hspace{2pt}\right) + \ln n\right] \ln n} \nonumber
\\  & = \sqrt{3K_n\ln n}  -
\sqrt{3\left[K_n  \hspace{-1.27pt}+ \hspace{-1.27pt}
\sqrt{\hspace{0.35pt}\ln n} \left( \sqrt{\hspace{0.35pt}\ln n}
\hspace{-1.27pt}- \hspace{-1.27pt} \sqrt{3K_n}\hspace{2pt}\right)
\right ] \hspace{-1.27pt} \ln n} \nonumber \\  & \geq \sqrt{3K_n\ln
n} - \sqrt{3K_n\ln n} \nonumber
\\  & =  0. \nonumber
\end{align}
From (\ref{pn7}) and condition $\frac{K_n}{P_n} = O\left(\frac{1}{n
\ln n}\right) $, it is clear to see (\ref{cp_c3}) due to
\begin{align}
p_n  & = \frac{K_n}{P_n} \cdot [1-o(1)]  = O\left(\frac{1}{n \ln
n}\right)  . \label{exprpn}
\end{align}
From (\ref{pn7}) and condition $\frac{{K_n}^2}{P_n} =
O\left(\frac{1}{ \ln n}\right)$, then (\ref{cp_c4}) is true
\begin{align}
{p_n}^2 P_n  & = \frac{{K_n}^2}{P_n} \cdot \big\{[1-o(1)]\big\}^2 =
O\left(\frac{1}{ \ln n}\right). \label{pn2Pn}
\end{align}
Below we will show (\ref{sn_prf}), where $s_n$ is specified in
(\ref{cp_c5}). Owing to ${p_n}^2 P_n \sim  \frac{{K_n}^2}{P_n}$
given in (\ref{pn2Pn}), we obtain from (\ref{cp_c5}) that
 \begin{align}
s_n & = \frac{{K_n}^2}{P_n} \cdot  [1-o(1)] \cdot \left[1 -
(n-2){p_n} + 2 {p_n} - \frac{{p_n}^2 P_n}{2} \right], \nonumber
\end{align}
which will result in (\ref{sn_prf}) once we derive
\begin{align}
 -(n-2) {p_n} - \frac{{p_n}^2 P_n}{2}  & =
 - O \left(\frac{1}{ \ln n}\right) . \label{n2pn}
\end{align}
(\ref{n2pn}) clearly follows from (\ref{exprpn}) and (\ref{pn2Pn}).

We have proved that (\ref{pnPn}) (\ref{Kngeq}) (\ref{cp_c3})
(\ref{cp_c4}) and (\ref{sn_prf}) all hold with $p_n$ and $s_n$ set
in (\ref{pn}) and (\ref{cp_c5}), respectively, provided the
conditions in Lemma \ref{cp_rig_er}. Then as noted before, with
$q_n$ set as $s_n$, we have established Lemma \ref{cp_rig_er} in
view of (\ref{cp_erg_rkg_rgg_ineq}). \qedb

\end{document}